\documentclass[compsoc, conference, a4paper, 10pt, times]{IEEEtran}
\pdfoutput=1 
\usepackage{cite}
\usepackage{amsmath,amssymb,amsfonts}
\usepackage{algorithmic}
\usepackage{graphicx}
\usepackage{textcomp}
\usepackage{xcolor}
\usepackage{booktabs}
\usepackage[hidelinks]{hyperref}
\usepackage{subcaption}
\usepackage{multirow}
\usepackage{stfloats}
\usepackage{todonotes}
\usepackage{paralist}

\usepackage{makecell}

\begin{document}

\title{Flashy Backdoor: Real-world Environment Backdoor Attack on SNNs with DVS Cameras}

  \author{
  \IEEEauthorblockN{Roberto Riaño}
    \IEEEauthorblockA{\textit{Radboud University, The Netherlands}\\
    \textit{Ikerlan Research Centre, Spain}\\
    roberto.rianohidalgo@ru.nl}
     \and
    \IEEEauthorblockN{Gorka Abad}
    \IEEEauthorblockA{\textit{Radboud University, The Netherlands}\\
    \textit{Ikerlan Research Centre, Spain}\\
    abad.gorka@ru.nl}
    \and
    \IEEEauthorblockN{Stjepan Picek}
    \IEEEauthorblockA{\textit{Radboud University, The Netherlands}\\
    stjepan.picek@ru.nl}
    \and
    \IEEEauthorblockN{Aitor Urbieta}
    \IEEEauthorblockA{\textit{Ikerlan Research Centre, Spain}\\
    aurbieta@ikerlan.es}
    }

\maketitle

\begin{abstract}
While security vulnerabilities in traditional Deep Neural Networks (DNNs) have been extensively studied, the susceptibility of Spiking Neural Networks (SNNs) to adversarial attacks remains mostly underexplored. Until now, the mechanisms to inject backdoors into SNN models have been limited to digital scenarios; thus, we present the first evaluation of backdoor attacks in real-world environments.

We begin by assessing the applicability of existing digital backdoor attacks and identifying their limitations for deployment in physical environments. To address each of the found limitations, we present three novel backdoor attack methods on SNNs, i.e., Framed, Strobing, and Flashy Backdoor. We also assess the effectiveness of traditional backdoor procedures and defenses adapted for SNNs, such as pruning, fine-tuning, and fine-pruning. The results show that while these procedures and defenses can mitigate some attacks, they often fail against stronger methods like Flashy Backdoor or sacrifice too much clean accuracy, rendering the models unusable.

Overall, all our methods can achieve up to a 100\% Attack Success Rate while maintaining high clean accuracy in every tested dataset. Additionally, we evaluate the stealthiness of the triggers with commonly used metrics, finding them highly stealthy. Thus, we propose new alternatives more suited for identifying poisoned samples in these scenarios. Our results show that further research is needed to ensure the security of SNN-based systems against backdoor attacks and their safe application in real-world scenarios. The code, experiments, and results are available in our repository.\footnote{\url{https://anonymous.4open.science/r/Flashy_backdoor/README.md}}

\end{abstract}



\section{Introduction}
Deep Neural Networks (DNNs) have revolutionized the field of Artificial Intelligence (AI), offering unprecedented advancements in various tasks, such as image recognition~\cite{Image-classification-cars-example}, natural language processing~\cite{NLP2}, or speech recognition~\cite{speech-recognition}. Their capacity to process and learn from vast amounts of data makes them ideal for various applications, ranging from medical analysis~\cite{medical-CNN} to autonomous driving systems~\cite{autopilot} and personal assistants~\cite{voiceAssistant}.

However, despite their success and wide application, DNN training is still a problem, as it demands significant computational resources and time, especially for large models like, e.g., GPT-3~\cite{GPT3_large}. Training such models requires extensive hardware and consumes a considerable amount of energy, posing environmental and economic challenges. For example, assuming that the GPT-3 model was trained using 1024 NVIDIA A100 GPUs in a data center, the time required to train this model would be 34 days~\cite{GPT3_time}, consuming about 235 MWh. Moreover, according to the method proposed by Lannelongue et al.~\cite{carbon}, it would have a carbon footprint of 62.47 T CO$_2$e (CO2 equivalent), which would take 5\,680 years\footnote{\url{https://calculator.green-algorithms.org/}} for a tree to process. This immense carbon footprint has led researchers to explore alternative, more sustainable models.

The motivation for using Spiking Neural Networks (SNNs) extends beyond their computational capabilities, emerging as a promising solution to this challenge~\cite{SNN, DL_SNN, training_SNN, enhance_SNN}. Unlike conventional DNNs that operate on continuous activations, SNNs simulate the spiking behavior observed in biological neurons~\cite{spiking_yolo}; this unique characteristic allows SNNs to capture the temporal dynamics of information processing, mirroring the precise timing and synchronization observed in the human brain~\cite{bioSpike}. SNNs exhibit a form of event-driven computation, meaning they only activate when necessary, leading to potential energy efficiency improvements compared to traditional neural network architectures. This advantage was demonstrated by Kundu et al.~\cite{eficient}, who showed that SNNs achieved up to a $12.2\times$ improvement in computational energy efficiency compared to DNNs with a similar parameter count. Additionally, SNNs can be more robust to noise and perturbations, allowing them to maintain accuracy in real-world situations under varying environmental conditions~\cite{SNN_backprop}.

The inherent ability of SNNs to manage temporal dependencies within data makes them particularly useful in tasks where the sequence and timing of events matter, such as in sensory processing~\cite{DL_SNN}. This capability becomes evident when employing data acquired through Dynamic Vision Sensor (DVS) cameras~\cite{DVS}. Unlike traditional cameras capturing absolute brightness at a constant rate, DVS cameras do not capture conventional video frames but instead record changes in light intensity at a pixel level, providing high temporal resolution and low-latency data~\cite{SNN_driving, Efficient_Perception}, referred to as neuromorphic data. This type of data aligns perfectly with the spike-based communication of SNNs, highlighting the interaction between the distinct features of both SNNs and neuromorphic data.

These capabilities are particularly useful in scenarios requiring rapid and efficient processing, such as autonomous driving~\cite{ SNN_driving, newAutoSNN} or robotics~\cite{DVS-drone}, where DVS cameras can work together with traditional sensors to capture the surrounding environment better to assess the current scenario. These cameras can also be used with SNNs to process the environment more efficiently than traditional cameras and DNNs. Event-based data is helpful when conventional data falls short~\cite{newAutoSNN, DVS-drone, Multivehicle, adverse_SNN} due to energy constraints, difficult lighting situations, or high-speed processing requirements.

Energy constraints are particularly critical in autonomous vehicles, where on-device processing is necessary to ensure real-time responsiveness and minimize dependence on cloud-based computation. For instance, a lightweight, fully unsupervised DNN adaptation algorithm for lane detection demonstrated real-time model adjustments, meeting similar constraints to those found in some existing level 3 autonomous cars~\cite{autonomousEnergy}.
However, such constraints will be harder to maintain with the increasing number of sensors and processing required, as autonomous vehicles typically utilize a vast array of sensors such as LiDAR, radar, and multiple cameras to perceive their environment, which significantly increases the data processing requirements and energy consumption~\cite{autonomousSensors}. SNNs paired with DVS cameras offer a more energy-efficient and low-latency alternative for scenarios where real-time processing is critical. Moreover, due to their event-driven nature and sparse activity, SNNs can drastically reduce power consumption and processing requirements while maintaining high computational efficiency and low latency. For instance, the Spiking Autonomous Driving~\cite{newAutoSNN} system integrates perception, prediction, and planning into a single end-to-end SNN model, showing competitive performance while significantly lowering energy consumption.
What is more, SNNs demonstrate broad applicability in other domains, including computer vision~\cite{action-recognition}, speech recognition~\cite{SNNspeech}, and medical diagnosis~\cite{stroke}. Despite the conventional belief that DNNs exhibit superior accuracy, recent findings indicate a diminishing or even disappearing performance gap compared to SNNs~\cite{DL_SNN}.

Although SNNs offer promising benefits and practical applications, a thorough assessment of their security aspects is necessary before they can be utilized in real-world situations. Many studies show ways in which a malicious actor could take advantage of insecure DNNs; examples include inference attacks~\cite{inference-survey}, evasion attacks~\cite{survey}, and poisoning attacks~\cite{poisoning-survey}. The consequences of these attacks can be dire, ranging from manipulated facial recognition systems and stealthy malware evasion~\cite{face_light, malware_evasion} to incorrect medical diagnosis~\cite{incorrect_medical_attack} or even tricking autonomous vehicles into misidentifying road signs or obstacles~\cite{adversarialSignsYoloV5}. However, the applicability of these attack methods to SNNs has yet to be studied.

This work contributes to evaluating SNN vulnerability by analyzing its susceptibility to backdoor attacks, emphasizing its suitability and relevance in real-world contexts. 
First, we test the backdoor attack presented by Abad et al.~\cite{paperGorka2}, the first to introduce backdoor attacks to SNNs within the digital domain. Their work demonstrated how hidden triggers embedded in the input data can cause misclassification in SNNs without degrading performance on clean data. Then, we propose two alternative trigger methods to solve the limitations we found when replicating these attacks for physical environments. We call these trigger methods 
\emph{Framed} and \emph{Strobing} triggers. 
Finally, after evaluating the results obtained by those methods, we generate our novel physical-environment backdoor, \emph{Flashy Backdoor}, evaluating its effectiveness in real-world scenarios. This highlights how vulnerabilities can be exploited in actual deployment settings, where inputs are sourced from the physical world rather than synthetic data or simulated scenarios.
To our knowledge, this paper presents the first backdoor attack on SNNs in real-world environments. Previously, all existing backdoor attack methods on SNNs were designed for digital environments, making them not applicable to physical scenarios. We demonstrate that carefully crafted triggers, specifically designed with the challenges and limitations of DVS cameras and neuromorphic data in mind, can be virtually modeled and successfully reproduced in physical environments., achieving up to a 100\% Attack Success Rate (ASR). Our contributions also include:
\begin{compactitem}
    \item Analyzing the suitability of previously proposed backdoor attacks for SNNs and introducing new attack variants that, while achieving at least 99\% ASR, do not show any degradation in clean accuracy.

    \item Testing procedures and defenses from the image domain to assess their efficacy on each proposed attack.

    \item Expanding the DVS128-Gesture dataset by adding new examples, including poisoned samples with physical triggers captured using a DVS camera, to test the attack's impact in real-world physical environments.
    
    \item Analyzing and comparing the stealthiness of our proposed methods with metrics such as the average Mean Square Error (MSE), Structural Similarity Index Metric (SSIM), average entropy analysis, and Peak Signal-to-Noise Ratio (PSNR). Finally, we explore other not-so-traditional methods to detect the triggers, like analyzing the number of activations per channel or the entropy distribution of the activation values per frame. 

\end{compactitem}

\section{Background}
\subsection{Spiking Neural Networks}

\begin{figure}[ht]
    \centering
    \includegraphics[width = 1\linewidth, trim= 35pt 0pt 35pt 0pt, clip]{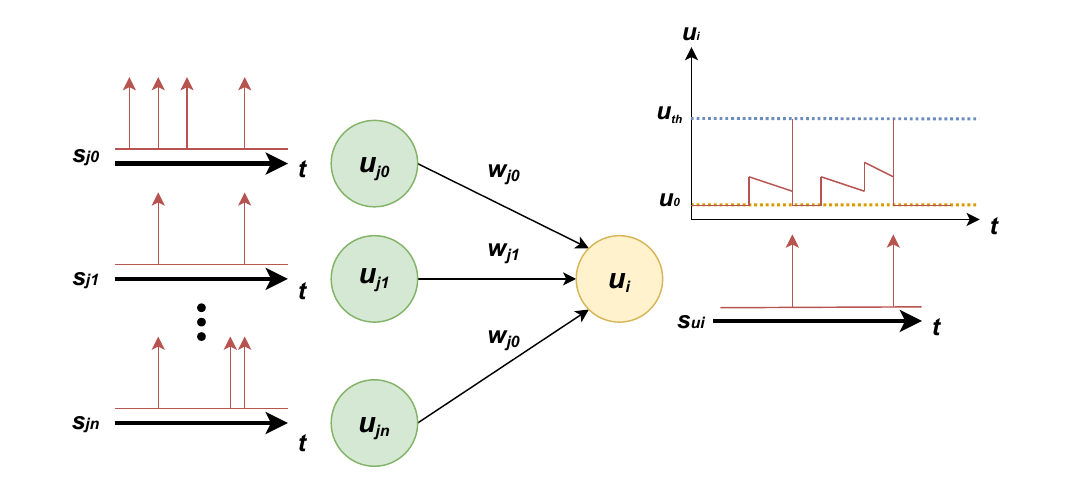}
    \caption{Workflow diagram of a LIF neuron, where $\mathbf{s}_j$ are the input spike trains, $\mathbf{u}_j$ are the membrane potentials of the presynaptic neurons, $\mathbf{u}_j$ are the synaptic weights, and $\mathbf{u}_i$ is the membrane potential of the neuron. When the membrane potential reaches the threshold $u_{th}$, the neuron fires an output spike $\mathbf{u}_{i}$.}
    \label{fig:SNN_model}
\end{figure}

SNNs, similar to traditional DNNs, are composed of layers of interconnected spiking neurons, including input, hidden, and output layers. However, unlike conventional DNNs that operate on continuous activations, SNNs simulate the spiking behavior observed in biological neurons~\cite{bioSpike} by using discrete events, known as spikes, to represent the output of a neuron. These spikes are generated when the neuron's membrane potential reaches a certain threshold (see Figure~\ref{fig:SNN_model}). This unique characteristic allows SNNs to capture the temporal dynamics of information processing, mirroring the precise timing and synchronization observed in the human brain~\cite{spiking_yolo}, while maintaining very low power consumption when paired with sparse input data~\cite{eficient}. Furthermore, recent studies have demonstrated the effectiveness of SNNs across different applications~\cite{SNN_prommise}, such as speech recognition, image classification, and robotics, further emphasizing their relevance in real-world contexts.

The Leaky Integrate-and-Fire (LIF)~\cite{LIF} model is a widely adopted method to model spiking neurons. This model represents the neuron's membrane potential over time, accumulating from its inputs. When the potential surpasses a certain threshold $u_{th}$, the neuron fires a spike and returns to its resting state. When the neuron does not receive any input, it loses its potential at a fixed rate until it either reaches the resting state or receives an input spike. The LIF model effectively captures the essential dynamics of spiking neurons by integrating incoming signals and incorporating mechanisms for both firing and potential decay, thereby mimicking the behavior observed in biological neurons~\cite{bioSpike}.
The behavior of each neuron in this model can be expressed with the following equations~\cite{SNN_formulation}:
\begin{equation}
\label{eq:LIF1}
u(t') = (1-s(t')) \cdot (\lambda u(t) + \sum_j w_j s_j(t')) +  s(t') \cdot u_0.
\end{equation}

\begin{equation}
\label{eq:LIF2}
 s(t') = \left\{
\begin{array}{l}

1 ,\quad  \text{if } u(t') \geq u_{th} \\
0 ,\quad \text{otherwise}
\end{array}
\right.
\end{equation}

Equation~\eqref{eq:LIF1} represents the temporal evolution of the neuron's membrane potential $u(t')$ before and after producing a spike $s$ at time $t'$. The first part of this equation is the evolution of the membrane before producing a spike, where $w_j$ are the synaptic weights (the parameter with which each incoming spike is multiplied), and $s_j(t')$ are the inputs to the neuron. The term $\lambda$ represents the leaky part of the LIF model; if $\lambda$ is set to 1, it would be equal to an Integrate and Fire (IF) neuron. However, when set lower than 1, it represents the rate at which the membrane potential will decrease toward its resting state in the absence of inputs. On the other hand, when the neuron produces a spike, the membrane potential is reset to its resting state $u_0$.

Equation~\eqref{eq:LIF2} represents the firing condition of the neuron, being $s(t')$ the output spike at time $t'$. This output will be 0 until the membrane potential reaches a threshold denoted by $u_{th}$. Then, at that time, a spike will be added to the output sequence of the neuron, known as spike train, resulting in the reset of the membrane potential in Equation~\eqref{eq:LIF1}.

In Equation~\eqref{eq:LIF2}, we can see one of the challenges in training SNNs; due to the non-differentiability of the spike activity, it is not possible to calculate the derivatives during backpropagation, and thus the direct use of traditional gradient-based optimization methods~\cite{training_SNN_gradients} is unfeasible. Various methods have been proposed to address this issue. Some involve training a traditional DNN to be converted into an SNN. This usually requires long simulations when applied to deep networks or more complicated datasets and results in longer inference times when compared to traditional DNNs~\cite{increased_activations}. Some common approaches are employed to reduce the time needed, like Robust Normalization, which increases the firing rates, or carefully reducing the firing thresholds of the spiking neurons. However, this also turns out in a lower power efficiency due to an increased amount of activations~\cite{increased_activations}. Nevertheless, this drawback can be highly mitigated with the compression methods proposed by Kundu et al.~\cite{eficient}.

Still, for high-performance SNN, direct training is crucial, i.e., training the SNN model from the ground up instead of converting the model from a DNN, especially for neuromorphic datasets~\cite{training_SNN_gradients}. For direct SNN training, there are two main options: Spike Timing Dependent Plasticity (STDP)~\cite{STDP} and surrogate gradiens~\cite{surrogate_gradients}. STDP updates the weights of the connections between neurons based on the timing of the spike propagation, considering two neurons, \emph{A} and \emph{B}, where \emph{A} precedes \emph{B} in the neural network architecture; if neuron \emph{A} fired before neuron \emph{B}, the connection between them is strengthened. Conversely, if neuron \emph{B} is fired before neuron \emph{A}, the connection between them is weakened.

On the other hand, surrogate training tackles the spikes' behavior, allowing the gradients to flow. During the training process, the surrogate gradients are used instead of the actual gradients of the loss function concerning the network parameters. These gradients are then used to update the weights using standard optimization algorithms like Stochastic Gradient Descent (SGD)~\cite{SGD}. There are several ways of computing the surrogate gradients; one of the first methods~\cite{surrogate_gradient_to_relu} consisted in approximating the derivative of a spiking neuron by the derivative of a truncated quadratic function, resulting in a ReLU function as the surrogate derivative. Other common surrogate functions include the fast sigmoid~\cite{fast_sigmoid} or the exponential function~\cite{exponential}; these functions mimic the behavior of spiking neurons while maintaining differentiability, allowing the gradients to be computed and backpropagated through the network during training.

\subsection{Backdoor Attacks}

Backdoor attacks focus on injecting hidden malicious behaviors into the models, behaving legitimately for benign inputs but abnormally in the presence of the predefined trigger~\cite{BadNets}. These attacks can be mostly classified into two main groups~\cite{minimal_rec}: poisoning-based and non-poisoning-based attacks. The former involves manipulating the training data to inject the desired behavior into the model. At the same time, Poisoning-based attacks can also be divided into two groups: clean-label attacks~\cite{clean_label_backdoor}, which maintain the appearance of the poisoned input consistent with their label, and dirty-label attacks~\cite{BadNets}, which change the poisoned examples' labels for them to fit the desired behavior. 
Non-poisoning-based attacks focus on embedding the backdoor into the model without altering the training data directly. These include weight-oriented backdoors~\cite{weight_backdoor}, structure-modified backdoors, also known as model injection~\cite{model_injection}, and code-poisoning backdoors~\cite{code_poisoning}.

There have been many studies on different types of triggers depending on the domain in which the backdoor is implemented. In the case of the image domain, a pixel pattern in a particular part of the image can serve as a trigger. Still, many other types have also been discussed in this domain; for example, Haoliang et al.~\cite{face_light} developed a novel color stripe pattern trigger, which could be projected onto real faces through a flickering LED to deceive face recognition models. In poisoning-based attacks, which we investigate, the models are trained on a mixture of clean and backdoor samples; that way, they learn only to misclassify the examples containing the trigger.

Further focusing on dirty label attacks, during the training process, a model $F_{\theta}(\mathbf{x})$ learns from a dataset comprising both clean examples $\mathcal{D}_{cl}$ of size $n$ and poisoned examples $\mathcal{D}_{ps}$ of size $m$. It is important to note that the size of $\mathcal{D}_{ps}$ is typically much smaller than $\mathcal{D}_{cl}$, with the poisoning rate indicating this proportion, expressed as $\varepsilon = \frac{m}{n}$. A clean sample is represented by $\{(\mathbf{x}, y)\} \in \mathcal{D}_{cl}$, where $\mathbf{x}$ is the input and $y$ is the corresponding label. A poisoned sample is represented by $\{(\hat{\mathbf{x}}, \hat{y})\} \in \mathcal{D}_{ps}$, where a mask $m$ is applied to the original input $\mathbf{x}$ to introduce a trigger $p$ at a specific location, resulting in the modified input $\hat{\mathbf{x}}$. In a dirty-label attack, the original label $y$ is replaced with the desired malicious label $\hat{y}$.
Equation~\eqref{eq:BD} shows the behavior of the backdoored model when encountering each type of sample: $\mathcal{R}$ represents the benign behavior and $\mathcal{B}$ the backdoor behavior.

\begin{equation}\label{eq:BD}
F_{\theta}(\mathbf{x})\left\{
\begin{array}{l}
\mathcal{R} \text{ if } \mathbf{x} \in {D}_{cl}\\
\mathcal{B} \text{ if } \mathbf{x} \in {D}_{ps}\\
\end{array}
\right.
\end{equation}

The main goal of the training process is to find the optimal parameters $\theta$ for a clean dataset that consists of $n$ samples. In this case, a second term is added to the equation to introduce the backdoor behavior to include the poisoned samples and labels~\cite{paperGorka2}.

\begin{equation}\label{eq:training_bd}
\begin{split}
  \theta' = & \arg\min_{\theta} \sum_{(\mathbf{x},y) \in {D}_{cl}}\mathcal{L}(F_{\theta}(\mathbf{x}), y)\; +\\
  & \sum_{(\hat{\mathbf{x}},\hat{y}) \in {D}_{ps}}\mathcal{L}(F_\theta (\hat{\mathbf{x}}),\hat{y})
\end{split}
\end{equation}
Following this methodology, we aim to maintain the correct behavior of the model when faced with clean inputs, represented by the first term of the equation, while ensuring the introduction of the backdoor behavior when faced with poisoned triggers, represented by the second term.

\section{Related Work}
Recent investigations undertook a comparative analysis of the inherent security vulnerabilities in SNNs, revealing their susceptibility to adversarial examples with both traditional data~\cite{adv_robustness_eval_noNeuro} and neuromorphic data from already existing datasets~\cite{adv_white_neuro}, data directly taken from DVS cameras~\cite{adv_white_neuroFromDVS}, or even using neuromorphic datasets with a model running on a neuromorphic chip in white-box scenarios~\cite{adv_white_neuroChip}. Researchers also tested the effect of adversarial examples on black-box scenarios with traditional data~\cite{SNN_secure, adv_black_noNeuro}, demonstrating that SNNs are susceptible to adversarial attacks with both traditional and neuromorphic data in white or black-box scenarios. Additionally, subsequent research~\cite{flip} has shown that SNNs are also prone to attacks hardware-level attacks, where intentional bit-flips can result in misclassification.

Despite these vulnerabilities, SNNs offer inherent advantages in terms of robustness due to their biological inspiration. Chen et al.~\cite{chen2024defending} proposed an SNN-based image purification model to defend against adversarial attacks. This method uses a noise extraction network and a non-blind denoising network to reconstruct clean images from noisy ones, enhancing the robustness of SNNs against adversarial examples. This approach leverages the visual masking effect and filtering theory to create a robust pre-processing module that can be integrated with various classifiers.
The Homeostatic Spiking Neural Networks model, developed by Geng and Li~\cite{HoSNNs}, introduces a threshold-adapting leaky integrate-and-fire neuron model. This model incorporates a self-stabilizing dynamic thresholding mechanism, clipping adversarial noise propagation and enhancing the robustness of SNNs in an unsupervised manner. The HoSNNs demonstrate significant improvements in accuracy against the Fast Gradient Sign Method and the Projected Gradient Descent attacks compared to traditional SNN models.

Other studies also considered the internal membrane perturbation in response to small disturbances. Din et al.~\cite{ding} introduced the concept of membrane potential perturbation dynamics, which analyzes the changes in a neuron's internal state when these disturbances are present. 
They used this concept to develop a dynamic LIF neuron with additional adjustable parameters, enabling better adaptation to input changes. 
Unlike previous methods, they focused on reducing the internal changes rather than just the spike outputs, demonstrating significant improvements in the robustness of SNNs against adversarial attacks.

To our knowledge, only a few works explore backdoor attacks in SNNs~\cite{paperGorka2, Paper_gorka_1, Time_Bandits, whispers, ANN_to_SNN_poisoning}, but none explore the possibility of real-world scenarios, focusing only on virtual environments. The first paper on backdoor attacks for SNNs~\cite{Paper_gorka_1} considered two types of triggers, a static and a moving square, present in all the frames of the poisoned data during training. However, this study only examined limited poisoning rates and trigger sizes without implementing any backdoor defense mechanisms.

The second one~\cite{paperGorka2} builds upon the initial study by exploring the feasibility of backdoor attacks in SNNs, with a particular focus on the stealthiness of the trigger and its robustness against defenses. The authors improved the performance of the static attacks proposed in~\cite{Paper_gorka_1} and developed two new trigger methods: smart triggers, which select the optimal trigger location and polarity, and dynamic triggers, which are highly stealthy. They evaluated the stealthiness of their newly proposed triggers using the SSIM metric, achieving up to 99.5\% SSIM, outperforming static and moving triggers at 98.5\% SSIM. Finally, they adapted image domain defenses for SNNs and neuromorphic data but found them ineffective against their backdoor attacks.

The two subsequent papers explore backdoor attacks on Federated Neuromorphic Learning (FedNL) scenarios. Abad et al. developed a novel backdoor attack strategy tailored to SNNs and FL, which distributes the backdoor trigger temporally and across malicious devices, enhancing the attack's effectiveness and stealthiness~\cite{Time_Bandits}. They evaluated the transferability of known FL attack methods to SNNs, finding suboptimal attack performance, and then explored single and multiple attackers to improve performance. Their approach achieved a 100\% attack success rate, highlighting the need for robust security measures in deploying SNNs in FL, particularly in the context of backdoor attacks.
Fu et al. presented an attack for FedNL-based systems using the concept of time division multiplexing, termed Spikewhisper~\cite{whispers}. This approach allows attackers to evade detection as multiple malicious clients can imperceptibly poison the model with different triggers at different timeslices. The stealthiness of Spikewhisper is derived from the time-domain divisibility of global triggers, with each malicious client injecting only one local trigger to a certain timeslice in the neuromorphic sample, configurable in both polarity and motion. Experiments on two different neuromorphic datasets demonstrated that the attack success rate of Spikewhisper is higher than temporally centralized attacks and is sensitive to the trigger duration.

Lastly, Jin et al.~\cite{ANN_to_SNN_poisoning} proposed a general backdoor attack framework based on data poisoning against SNNs trained by supervised learning methods. They analyzed the robustness differences between different learning methods and concluded that SNNs do not inherently possess robustness against backdoor attacks. They also showed that backdoors can be transferred from ANNs to SNNs during conversion, with a migration rate exceeding 99\%. 


\section{Physical Dataset Expansion}
\label{datasetExpansion}
In this section, we follow the process used to capture and adapt new dataset samples to the original DVS128-Gestures dataset~\cite{gestures}, along with the challenges of replicating digital triggers in physical environments. We chose this dataset as it is, to our knowledge, one of the few 
widely available options that provide data directly captured by DVS cameras for classification tasks rather than being converted from frame-based datasets. We also considered the ASL-DVS~\cite{ASL} dataset of American Sign Language gestures but discarded it due to the short time duration of the samples, i.e., 0.1 seconds, making it unsuitable for trigger insertion. The DVS128-Gestures dataset also allowed us to directly compare the performance of our attacks in digital settings with previously studied backdoor methods while being able to analyze the differences between the efficacy of the digital triggers and their physical reproductions.


\subsection{Dynamic Vision Sensors}
DVS cameras operate differently from traditional frame-based cameras, capturing asynchronous changes in light intensity (events) at a pixel level rather than producing full video frames~\cite{Gallego}.
Each pixel in a DVS camera independently reports changes in brightness, generating spikes of information when such changes occur.

Each event recorded by a DVS includes specific data: the ${X, Y}$ coordinates of the pixel location, the polarity $p$ indicating whether the change is an increase (\emph{ON} event) or decrease (\emph{OFF} event) in light intensity, and a timestamp $T$ that provides the precise time of the event, usually measured in microseconds. This efficient encoding scheme helps reduce energy consumption and data bandwidth~\cite{eficient}.

\begin{figure}[ht]
    \centering
    \includegraphics[width = 0.75\linewidth]{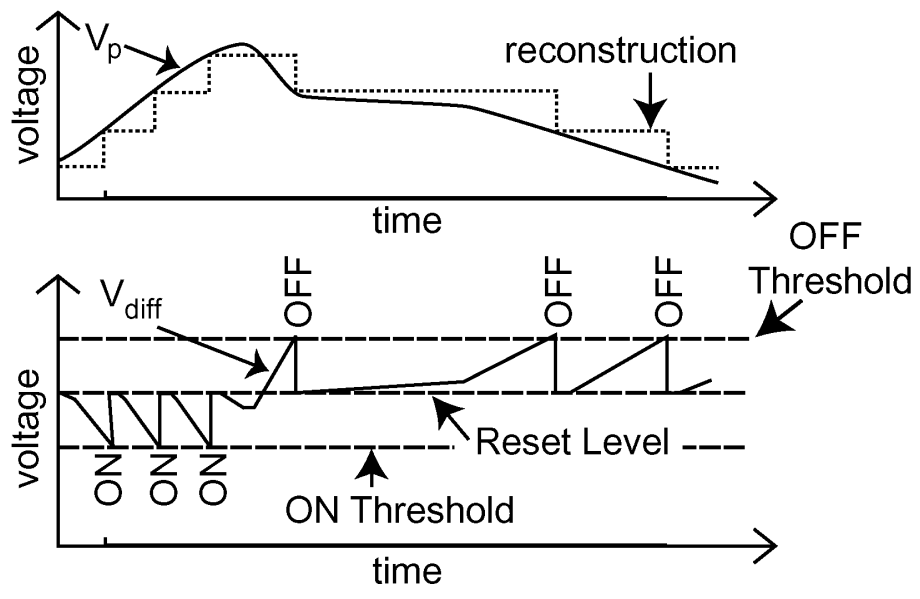}
    \caption{Principle of operation of a DVS described by Lichtsteiner et  al.~\cite{Lichtsteiner}.}
    \label{fig:DVSwork}
\end{figure}

The pixel operation in a DVS involves three main components~\cite{Lichtsteiner}: a photoreceptor, a differencing circuit, and a comparator. The photoreceptor measures light intensity and converts it into a logarithmic signal. The differencing circuit compares the current intensity to a stored reference value. When the difference exceeds a predefined threshold, the comparator generates an event (see Figure~\ref{fig:DVSwork}). 

\subsection{Data Collection Process}
The data collection process involved recording gestures performed by subjects under controlled conditions to match the settings of the original dataset using the Metavision EVK3–HD camera. The participants performed the same set of gestures as in the  DVS128-Gesture dataset.

We carefully adjust the camera parameters to align the captured data with the characteristics of the original dataset. Two of the most important parameters to obtain low noise, clear samples are \textit{Event Generation Thresholds}, which controls the thresholds that determine the minimum change in light intensity required for a pixel to generate an event, and \textit{Pixel Rest Time}, also known as the refractory period, which defines the minimum time between successive events from the same pixel. The tuning of these parameters is important to match the number of events present in the original samples while reducing background noise. For the specific parameter values, refer to Appendix~\ref{appendixC}. 

After capturing the new samples, we preprocess them to align with the data on the original dataset. As the Metavision EVK3–HD has a much higher resolution of 1280$\times$720, we crop our samples to a size of 512$\times$512 before downscaling them to the intended resolution of 128$\times$128, matching the DVS128-Gesture dataset samples' size. We also ensure the temporal resolution is aligned by truncating our samples to the intended longitude, each lasting about 6 seconds like the original dataset~\cite{gestures}.

\begin{figure}[ht]
    \centering
    \begin{subfigure}[b]{0.22\linewidth}
        \includegraphics[width=\linewidth]{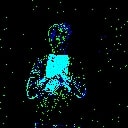}
    \end{subfigure}
    \hfill
    \begin{subfigure}[b]{0.22\linewidth}
        \includegraphics[width=\linewidth]{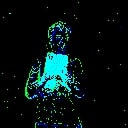}
    \end{subfigure}
    \hfill
    \begin{subfigure}[b]{0.22\linewidth}
        \includegraphics[width=\linewidth]{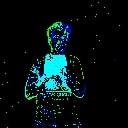}
    \end{subfigure}
    \hfill
    \begin{subfigure}[b]{0.22\linewidth}
        \includegraphics[width=\linewidth]{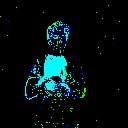}
    \end{subfigure}
\caption{Sample of a subject performing a clapping motion captured with our DVS camera.}
    \label{fig:CleanFrames3}
\end{figure}

\subsection{Trigger Reproduction and Challenges}

After assessing the static backdoor proposed by Abad et al.~\cite{Paper_gorka_1}. We observed two main challenges for applying this approach in physical environments: 1) the trigger is inserted in all frames of each poisoned sample, and 2) the trigger polarity and size remain fixed through all the sample frames in each experiment.
These limitations make the reproduction of triggers challenging due to the adaptive behavior of the DVS pixels, which adjust their firing thresholds based on recent light intensity changes. Unlike digital settings where triggers can be precisely inserted into specific positions and timing,
physical reproduction requires manipulating light sources while accounting for the DVS's sensitivity and adaptation mechanisms. A static trigger, such as a fixed visual pattern, would only produce events initially and then become undetectable as the pixels adapt. To be detected again while maintaining the same polarity would require continuously increasing or decreasing its luminosity. This approach is impractical in real-world scenarios, as such constant change would affect the surrounding area, introducing unwanted noise.





\section{Attacks on SNNs}
\label{Attacks}
\subsection{Threat Model}

We consider a scenario where a user wishes to train an SNN, $F$, but either lacks the resources to train the model independently or does not have the means to obtain its own dataset $D$.

The attacker's goal is to ensure that the backdoored model, $F_\theta$, behaves normally on clean data but follows the attacker's desired outcome when a specific trigger is present. To achieve this, we assume they have similar capabilities to prior research on backdoor injection~\cite{ssba, physicalBackdoor, invisible_pert, clean_label_backdoor}. They can inject a small number of poisoned samples into the training data, $D_{train}$, either by tampering with the public dataset the user obtains or by manipulating the data during outsourced training through an MLaaS provider (e.g., Google Cloud, AWS). Limiting the amount of poisoned data is essential to avoid detection during any dataset analysis, as introducing a large number of poisoned samples increases the risk of being identified. Additionally, the attacker has no further control over the model training, knowledge of the internal weights, or the architecture of the trained model.

In either scenario, the user holds back a validation set, $D_{valid}$, to check the model's performance, which the attacker cannot access.

 Moreover, no further assumptions need to be made on either scenario to perform our attacks in the physical setting. We only assume that the attacker can mimic the digitally crafted triggers and access the camera used by the user to capture the physical samples to perform the attack.

\subsection{Methodology}
To ensure the reliability of our results, we conducted all the digital environment experiments on three datasets, NMnist~\cite{N-MNIST}, Cifar10-DVS~\cite{CIFAR_DVS}, and DVS128-Gesture~\cite{gestures}, and repeated each experiment several times to minimize variability.
We chose the NMnist, Cifar10-DVS, and DVS128-Gesture datasets for our experiments due to their widespread use in neuromorphic research~\cite{adv_white_neuro, ANN_to_SNN_poisoning, paperGorka2, whispers} (for a more detailed description of each dataset or the models used, refer to Appendix~\ref{appendixB}); each one of them was preprocessed to fit the input requirements of our SNN models. These datasets provide a variety of challenges and are suitable for testing the robustness and effectiveness of backdoor attacks on SNNs. 

With the objective of overcoming the challenges discussed in Section~\ref{datasetExpansion},
we propose two novel attacks: \emph{Framed Backdoor} and \emph{Strobing backdoor}, which can be combined into the \emph{Flashy backdoor} to be applicable in real-world scenarios with DVS cameras.

\subsubsection{Framed Backdoor}
\begin{figure}[ht]
    \centering
    \begin{subfigure}{0.22\linewidth}
        \includegraphics[width=\linewidth]{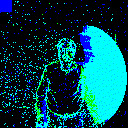}
        \caption{Frame 1}
    \end{subfigure}
    \hfill
    \begin{subfigure}{0.22\linewidth}
        \includegraphics[width=\linewidth]{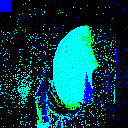}
        \caption{Frame 2}
    \end{subfigure}
    \hfill
    \begin{subfigure}{0.22\linewidth}
        \includegraphics[width=\linewidth]{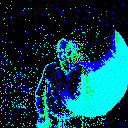}
        \caption{Frame 3}
    \end{subfigure}
    \hfill
    \begin{subfigure}{0.22\linewidth}
        \includegraphics[width=\linewidth]{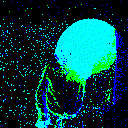}
        \caption{Frame 4}
    \end{subfigure}
    \caption{Example of a Framed trigger with a size of 10\% of the total frame inserted in the top-left position of the first 2 frames.}
    \label{fig:frame}
\end{figure}

We use this attack to evaluate the effect that the trigger size, number of frames, and the temporal location of the trigger have on the performance of the original~\cite{Paper_gorka_1} approach. Our experiments focus on a single position of the input space, and we vary the trigger's size, length, and temporal position for each possible polarity. We conduct our experiments with a $\varepsilon$ value of $0.1$ and the trigger's location set to the top-left (see Figure~\ref{fig:frame}). We experiment with four different trigger sizes, ranging from 10\% to 100\% of the original frame, and varying trigger lengths, ranging from a single frame to eight consecutive frames (of a total of sixteen). Lastly, we test the influence of temporal position by injecting the backdoor at the input space's start, middle, and end.

The function $\mathcal{A}(\mathbf{x},p,l,s,i,d)$ is a formal representation of the process of generating a set of poisoned inputs $\mathcal{D}_{ps}: \hat{\mathbf{x}}\in\mathcal{D}_{ps}$, where $\mathbf{x}$ denotes the clean sample that is subject to poisoning. The function takes in five parameters: $p$, representing the polarity of the trigger to be introduced, with four different options (ranging from $p=0$ to $p=3$, being these all the possible polarities on a pixel for any sample)
;$l$, representing the spacial location the trigger will be introduced in (e.g., top-right, middle, bottom-left);
and, $s$, representing the trigger size, or coverage in a frame, as a percentage of the input
. As the input samples are divided into $T$ frames, we use $i$ to select the initial frame where the insertion of the trigger will start, ranging from $0$ to $T-1$; and $d$, to define the duration of the trigger to be introduced, measured in frames. 

To select the frames to be poisoned in each sample, we use the following criteria:

$$
   \forall t \in [0,T ), \;\hat{\mathbf{x}}_t=\left \{
    \begin{array}{l}
        \hat{\mathbf{x}}_t,\;\; \text{ if }\;  t\in[i,i+d )\\
        \mathbf{x}_t,\;\; \text{ otherwise} 
    \end{array}
    \right.
$$
where $\hat{\mathbf{x}}_t$ represents the frames where the trigger is inserted and $\mathbf{x}_t$ represents the clean frames.

\subsubsection{Strobing Backdoor}

\begin{figure}[ht]
    \centering
    \includegraphics[width=0.75\linewidth]{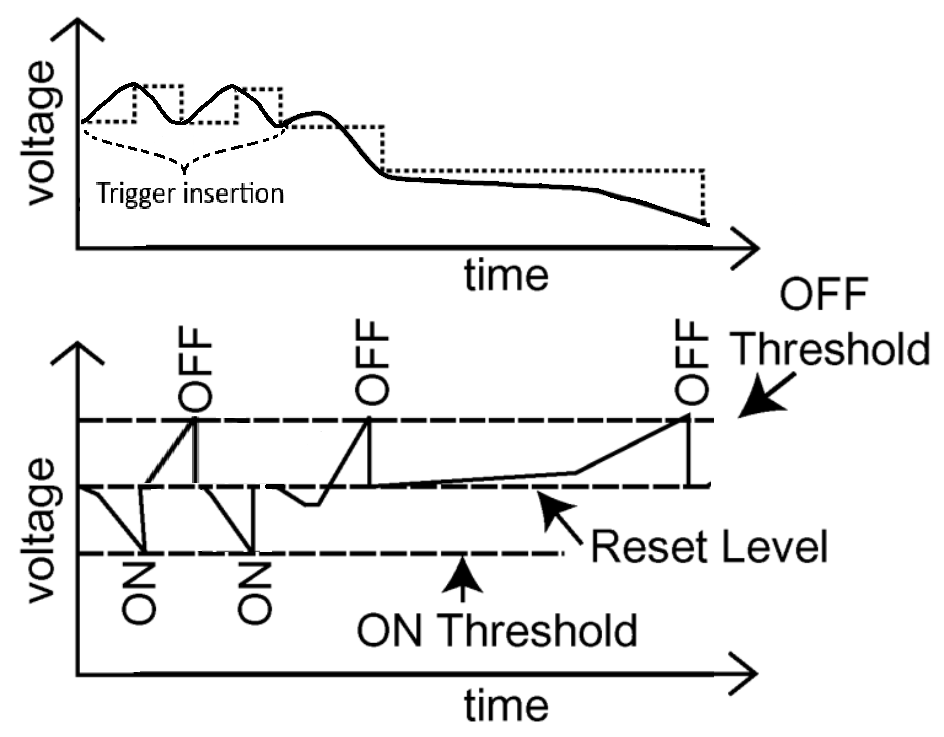}
    \caption{Principle of operation of the pixels of the DVS camera in the poisoned section where the trigger is inserted in the three first frames, leaving a gap between them to let the threshold reset.}
    \label{fig:PoisonframeDVS}
\end{figure}

\begin{figure}[ht]
    \centering
    \begin{subfigure}{0.22\linewidth}
        \includegraphics[width=\linewidth]{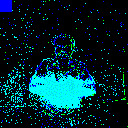}
        \caption{Frame 1}
    \end{subfigure}
    \hfill
    \begin{subfigure}{0.22\linewidth}
        \includegraphics[width=\linewidth]{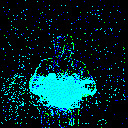}
        \caption{Frame 2}
    \end{subfigure}
    \hfill
    \begin{subfigure}{0.22\linewidth}
        \includegraphics[width=\linewidth]{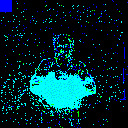}
        \caption{Frame 3}
    \end{subfigure}
    \hfill
    \begin{subfigure}{0.22\linewidth}
        \includegraphics[width=\linewidth]{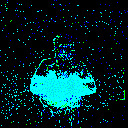}
        \caption{Frame 4}
    \end{subfigure}
    \caption{Example of a Strobing trigger with a size of 10\% of the total frame inserted in the top-left position of the first 2 frames leaving a clean gap of 1 frame between them.}
    \label{fig:strobe}
\end{figure}

This attack implements the idea of an intermittent trigger to exploit the temporal dynamics of SNNs and evade the adaptative threshold of DVS cameras. The DVS adapts each pixel's firing threshold to the luminosity of the last event they generated; consequently, if a static trigger is introduced, it would only fire once before adapting and would not be captured again until a change is detected. Thus, we approach the \emph{Strobing Backdoor} using the same methodology as the \emph{Framed Backdoor}, but this time, we inserted a certain amount of clean frames between each pair of poisoned frames. This change allows the DVS to adapt to the environment and reset its luminosity threshold (see Figure~\ref{fig:PoisonframeDVS}), enabling the attacker to maintain the same trigger luminosity, shape, and position while making sure that the trigger appears in the intended frames (see Figure~\ref{fig:strobe}). We also evaluate the effect that the size of the clean gap between the trigger frames has on the training process and the model's performance.

Formally, for a given polarity $p$, a spacial location $l$, trigger size $s$, and clean gap $g$, the trigger disappears and reappears on the same location with frequency $g$ on $d$ consecutive frames $f$, starting on the $i^\text{th}$ frame and with a duration of $d$ frames. The backdoor function $\mathcal{A}(\mathbf{x},p,l,s,i,d,g)$ takes the parameters described to create a poisoned set of inputs $\mathcal{D}_{ps}$. To select the poisoned frames of each sample, we use the following formulation:

$$
   \forall t \in [0,T ), \;\hat{\mathbf{x}}_t=\left \{
    \begin{array}{l}
        \begin{split}
        \hat{\mathbf{x}}_t,\;\; \text{ if }\; & t\in[i,i+d ) \;\; \wedge \;\; (t-i) \\& \equiv 0 \bmod(g+1)
        \end{split}\\
        \mathbf{x}_t,\;\; \text{ otherwise} 
    \end{array}
    \right.
$$

where $\hat{\mathbf{x}}_t$ represent sthe frames where the trigger is inserted and $\mathbf{x}_t$ the clean frames. Note that if $g=0$, this would be equivalent to the Framed Backdoor.

\subsubsection{Flashy Backdoor}
\begin{figure}[ht]
    \centering
    \begin{subfigure}{0.22\linewidth}
        \includegraphics[width=\linewidth]{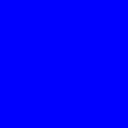}
        \caption{Frame 1}
    \end{subfigure}
    \hfill
    \begin{subfigure}{0.22\linewidth}
        \includegraphics[width=\linewidth]{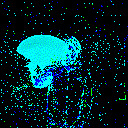}
        \caption{Frame 2}
    \end{subfigure}
    \hfill
    \begin{subfigure}{0.22\linewidth}
        \includegraphics[width=\linewidth]{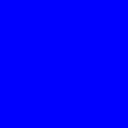}
        \caption{Frame 3}
    \end{subfigure}
    \hfill
    \begin{subfigure}{0.22\linewidth}
        \includegraphics[width=\linewidth]{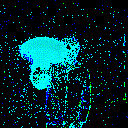}
        \caption{Frame 4}
    \end{subfigure}
    \caption{Example of a Flash trigger inserted in the first 2 frames leaving a clean gap of 1 frame between them.}
    \label{fig:flash}
\end{figure}

Finally, we combine the ideas of both \emph{Framed} and \emph{Strobing Backdoors} to create the \emph{Flashy backdoor} and overcome the limitations found for real-world environment attack deployment. The main idea is to simulate the light of a strobing flashlight being pointed at the camera as our trigger while ensuring that the DVS camera would have enough time to reset its luminosity threshold, following the same principle explained in Figure~\ref{fig:PoisonframeDVS} but expanded to all the pixels of the DVS camera instead of a specific section (see Figure~\ref{fig:flash}). By doing this, we can maintain the same luminosity and trigger size for all our triggers.

\subsubsection{Physical Trigger Reproduction}
Once we tested this premise in virtual environments, we moved on to a physical environment. 


We train a backdoored model with virtually introduced triggers on the DVS128-Gesture. We then capture new samples replicating the digital trigger by intermittently shining two common light sources, i.e., a household flashlight (75-110 lumens) and a regular red laser pointer directed at the camera. It is important to note that the attacker does not introduce any new samples on the dataset or modify the already trained networks in any way, i.e., no fine-tuning with the new poisoned samples was performed. This way, we evaluate the ability of the attacker to recreate the virtual triggers in physical environments.



\begin{figure}[ht]
    \centering
    \begin{subfigure}[b]{0.22\linewidth}
        \includegraphics[width=\linewidth]{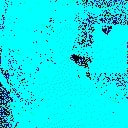}
        \caption{L.frame 1}
    \end{subfigure}
    \hfill
    \begin{subfigure}[b]{0.22\linewidth}
        \includegraphics[width=\linewidth]{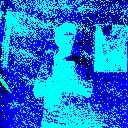}
        \caption{L.frame 2}
    \end{subfigure}
    \hfill
    \begin{subfigure}[b]{0.22\linewidth}
        \includegraphics[width=\linewidth]{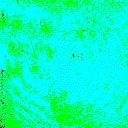}
        \caption{L.frame 3}
    \end{subfigure}
    \hfill
    \begin{subfigure}[b]{0.22\linewidth}
        \includegraphics[width=\linewidth]{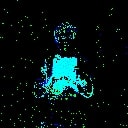}
        \caption{L.frame 4}
    \end{subfigure}

    \begin{subfigure}[b]{0.22\linewidth}
        \includegraphics[width=\linewidth]{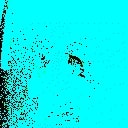}
        \caption{F.frame 1}
    \end{subfigure}
    \hfill
    \begin{subfigure}[b]{0.22\linewidth}
        \includegraphics[width=\linewidth]{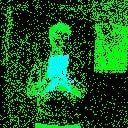}
        \caption{F.frame 2}
    \end{subfigure}
    \hfill
    \begin{subfigure}[b]{0.22\linewidth}
        \includegraphics[width=\linewidth]{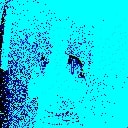}
        \caption{F.frame 3}
    \end{subfigure}
    \hfill
    \begin{subfigure}[b]{0.22\linewidth}
        \includegraphics[width=\linewidth]{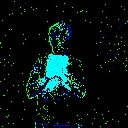}
        \caption{F.frame 4}
    \end{subfigure}
    
    \caption{First four collected frames of samples containing triggers. In frames (a)-(d), the trigger is inserted with the laser pointer; in frames (e)-(h), the trigger is inserted with the flashlight.}
    \label{fig:PoisonedFrames3}
\end{figure}

To replicate the digital triggers with our light sources, we set the flashlight at a distance of 20 cm from the sensor with an angle of 45\textdegree~from the center of the lens. This distance was chosen to take into account the power and directionality of the flashlight, as we noted that increasing the distance between the sensor and the flashlight resulted in the light beam spreading out and losing the capacity to introduce the trigger on the samples correctly. This could be solved using a more powerful flashlight or a more directional light beam. That did not pose a problem when using the laser pointer, as the light beam of a laser is highly directional and does not spread or lose intensity at reasonable distances, so we flashed the camera from different positions and distances with no difference in the resulting trigger.
(see Figure~\ref{fig:PoisonedFrames3}).

\section{Evaluation}

\subsection{Experimental Results}
We evaluate the attacks with commonly used metrics:
\begin{compactenum}
    \item ASR measures the backdoor performance of the model based on a holdout fully backdoored dataset. To ensure accurate measurement, for every poisoned sample $\hat{\textbf{x}}$, we verify that the original label $y$ differs from the target backdoor label $\hat{y}$. If there is any sample $\hat{\textbf{x}} \in {D}_{ps}$ where $y = \hat{y}$, we replace that sample with another one to guarantee the correct assessment of ASR.
    \item Clean accuracy measures the model's performance on a holdout clean dataset.
\end{compactenum}

\begin{table}[t]
\centering
\caption{Comparison of the results with different trigger sizes of $p=1$ inserted at the start in the DVS128-Gesture dataset.}\label{tab:1}
\begin{tabular}{@{}cccc@{}}
\toprule
\textbf{ASR\%} & \textbf{Clean ACC\%} & \textbf{Trigger length} & \textbf{Trigger size\%} \\ \midrule
12.38 $\pm$ 0.05 & 90.16 $\pm$ 0.02 & 5 & 10 \\
63.43 $\pm$ 41.36 & 89.0 $\pm$ 0.08 & 6 & 10 \\
15.51 $\pm$ 0.22 & 90.51 $\pm$ 0.05 & 7 & 10 \\
58.68 $\pm$ 26.12 & 90.28 $\pm$ 0.03 & 8 & 10 \\ \midrule
13.54 $\pm$ 0.19 & 89.58 $\pm$ 0.04 & 3 & 20 \\
60.53 $\pm$ 37.85 & 89.12 $\pm$ 0.15 & 4 & 20 \\
71.88 $\pm$ 44.58 & 88.89 $\pm$ 0.07 & 5 & 20 \\
75.46 $\pm$ 36.12 & 90.74 $\pm$ 0.06 & 6 & 20 \\ \midrule
13.54 $\pm$ 0.09 & 88.43 $\pm$ 0.06 & 2 & 30 \\
40.97 $\pm$ 51.67 & 89.35 $\pm$ 0.13 & 3 & 30 \\
88.77 $\pm$ 7.56 & 90.39 $\pm$ 0.01 & 4 & 30 \\
100 $\pm$ 0.0 & 91.2 $\pm$ 0.02 & 7 & 30 \\ \midrule
19.33 $\pm$ 0.53 & 88.89 $\pm$ 0.05 & 1 & 100 \\
98.84 $\pm$ 0.04 & 90.62 $\pm$ 0.0 & 2 & 100 \\
100 $\pm$ 0.0 & 90.62 $\pm$ 0.03 & 3 & 100 \\
100 $\pm$0.0 & 92.82 $\pm$0.03 & 4 & 100 \\ \bottomrule
\end{tabular}

\end{table}

All attacks were tested on each dataset three times, and the standard deviation (STD) was measured to account for variability.

\begin{table}[b]
\centering
\caption{Comparison of the results with different temporal locations of the trigger with $p=1$ and size of 100\% of the original frame on DVS128-Gesture dataset}\label{tab:2}
\begin{tabular}{@{}cccc@{}}
\toprule
\textbf{ASR\%} & \textbf{Clean ACC \%} & \multicolumn{1}{c}{\textbf{Trigger length}} & \textbf{Position} \\ \midrule
19.33 $\pm$ 0.53 & 88.89 $\pm$ 0.05 & 1 & Start \\
98.84 $\pm$ 0.04 & 90.62 $\pm$ 0.0 & 2 & Start \\
100 $\pm$ 0.0 & 90.62 $\pm$ 0.03 & 3 & Start \\ \midrule
47.34 $\pm$ 19.35 & 90.05 $\pm$ 0.01 & 1 & Middle \\
99.54 $\pm$ 0.01 & 91.67 $\pm$ 0.0 & 2 & Middle \\
100 $\pm$ 0.0 & 92.59 $\pm$ 0.0 & 3 & Middle \\ \midrule
11.11 $\pm$ 0.01 & 88.77 $\pm$ 0.01 & 1 & End \\
24.65 $\pm$ 0.25 & 90.74 $\pm$ 0.01 & 6 & End \\
86.0 $\pm$ 6.15 & 91.55 $\pm$ 0.05 & 8 & End \\ \bottomrule
\end{tabular}
\end{table}

\subsection{Framed Backdoor}

To assess the viability of the proposed attack, we first experiment with the BadNets~\cite{BadNets} approach, but limiting the position of the trigger to the upper-left corner and maintaining a constant poisoning rate $\epsilon$ of $0.1$. Then, we introduced the trigger in different time sections of the samples (start, middle, or end), with trigger duration ranging from 1 to 8 frames and four different trigger sizes: 10\%, 20\%, 30\%, and 100\% of the original image. The previously mentioned limitations allowed us to carefully examine the effect these parameters have on the resulting ASR as well as the clean accuracy of the model.

Our experiments reveal that the trigger's size is inversely correlated with the number of frames required to perform the backdoor effectively. When the trigger is the largest and presented at the start or middle, the attack achieved a 100\% ASR in all tested datasets for every possible polarity. Furthermore, in the most efficient polarities, it can be observed that the system is capable of achieving 100\% ASR accuracy within a mere 2 to 3 frames. Increasing the trigger's size also increased the stability of the results (see Table~\ref{tab:1}). Notably, every attack we tested maintained the same clean accuracy for all cases studied. We also observe that the temporal position affects the attack's effectiveness, performing worse the later it is introduced (see Table~\ref{tab:2}). We hypothesize that this effect is due to how spiking neurons operate, as they accumulate membrane potential over time as they receive input spikes. Introducing the trigger later in the sequence allows less time for the membrane potential to build up and reach the firing threshold, giving the trigger fewer opportunities to take effect. However, further testing is required to determine the exact cause of this.

\begin{table}[ht]
\centering
\caption{Comparison of the results with different trigger sizes of $p=1$ inserted at the start in the DVS128-Gesture dataset with a gap of 1 clean frame between each 2 poisoned frames.}\label{tab:3}
\begin{tabular}{@{}cccc@{}}
\toprule
\textbf{ASR\%} & \textbf{Clean ACC \%} & \multicolumn{1}{c}{\textbf{P frames}} & \multicolumn{1}{c}{\textbf{Trigger Size \%}} \\ \midrule
10.76 $\pm$ 0.02 & 89.35 $\pm$ 0.06 & 3 & 10 \\
11.11 $\pm$ 0.01 & 88.54 $\pm$ 0.07 & 4 & 10 \\
11.23 $\pm$ 0.02 & 88.77 $\pm$ 0.07 & 5 & 10 \\
12.15 $\pm$ 0.07 & 88.19 $\pm$ 0.07 & 6 & 10 \\ \midrule
12.15 $\pm$ 0.14 & 89.12 $\pm$ 0.1 & 3 & 20 \\
50.69 $\pm$ 40.72 & 90.16 $\pm$ 0.11 & 4 & 20 \\
51.27 $\pm$ 39.77 & 89.12 $\pm$ 0.07 & 5 & 20 \\
70.37 $\pm$ 52.67 & 90.86 $\pm$ 0.03 & 6 & 20 \\ \midrule
12.85 $\pm$ 0.07 & 89.00 $\pm$ 0.18 & 2 & 30 \\
71.53 $\pm$ 48.64 & 91.32 $\pm$ 0.09 & 5 & 30 \\
99.88 $\pm$ 0.0 & 91.44 $\pm$ 0.01 & 6 & 30 \\ \midrule
19.33 $\pm$ 0.53 & 88.89 $\pm$ 0.05 & 1 & 100 \\
100 $\pm$ 0.0 & 91.67 $\pm$ 0.06 & 2 & 100 \\
100 $\pm$ 0.0 & 91.44 $\pm$ 0.03 & 3 & 100 \\
100 $\pm$ 0.0 & 92.25 $\pm$ 0.02 & 4 & 100 \\ \bottomrule
\end{tabular}
\end{table}

\begin{table}[ht]
\centering
\caption{Comparison of the results with different temporal locations of the trigger with $p=1$ and a size of 100\% of the original frame on the DVS128-Gesture dataset with a gap of 1 clean frame between each 2 poisoned frames.}\label{tab:4}
\begin{tabular}{@{}cccc@{}}
\toprule
\textbf{ASR\%} & \textbf{Clean ACC \%} & \textbf{P frames} & \textbf{Position} \\ \midrule
19.33 $\pm$ 0.53 & 88.89 $\pm$ 0.05 & 1 & Start \\
100 $\pm$ 0.0 & 91.67 $\pm$ 0.06 & 2 & Start \\
100 $\pm$ 0.0 & 91.44 $\pm$ 0.03 & 3 & Start \\ \midrule
47.34 $\pm$ 19.35 & 90.05 $\pm$ 0.01 & 1 & Middle \\
99.88 $\pm$ 0.0 & 91.90 $\pm$ 0.01 & 2 & Middle \\
100 $\pm$ 0.0 & 92.13 $\pm$ 0.02 & 3 & Middle \\ \midrule
11.11 $\pm$ 0.01 & 88.77 $\pm$ 0.01 & 1 & End \\
50.93 $\pm$ 0.34 & 90.62 $\pm$ 0.01 & 4 & End \\
100 $\pm$ 0.0 & 91.20 $\pm$ 0.02 & 5 & End \\ \bottomrule
\end{tabular}
\end{table}

\subsection{Strobing Backdoor}

The next step is to analyze the effect a non-continuous trigger has on the model performance. We maintain the same limitations as in the previous test, but this time, we introduce a clean frame between each pair of poisoned frames. We proceed again to introduce the trigger in each time section, with trigger duration ranging from 3 to 13 frames.

Our testing shows that the attack's performance is mostly equal or occasionally better when the trigger is positioned at the start or middle (see Table~\ref{tab:3}), without observing any degradation in the clean accuracy (see Table~\ref{tab:3}). However, when positioned at the end, its performance increases compared to the \emph{Framed Backdoor} (see Table~\ref{tab:4}).
This approach also enhances the attack's stealthiness, as it requires the trigger to be present in fewer frames than its continuous counterpart, and it yields improved performance in cases where the trigger is not present at the start (see Table~\ref{tab:3}).

After doing a hyperparameter analysis, we conclude that modifying the trigger gap does not have any substantial effect on the clean accuracy. However, it positively affects ASR when the trigger is positioned at the end. Still, careful consideration of the gap is needed, as it can also make the training less stable for the backdoor behavior, and significant gaps can have the opposite effect on ASR (see Table~\ref{tab:gap_analisis}).

\begin{table}[ht]
\centering
\caption{Results of the hyperparameter analysis of the gap length in the DVS128-Gesture dataset, with a gap size ranging from 1 to 4 and the trigger introduced at the Start, Middle, or End of the poisoned samples with 3 poisoned frames with triggers of size 20\% on each one.}
\label{tab:gap_analisis}
\begin{tabular}{@{}cccc@{}}
\toprule
\textbf{ASR\%} & \textbf{Clean ACC \%} & \textbf{Gap size} & \textbf{Position} \\ \midrule
98.95 $\pm$ 0.35 & 90.27 $\pm$ 0.006 & 1 & START \\
96.41 $\pm$ 2.56 & 91.66 $\pm$ 0.015 & 2 & START \\
69.44 $\pm$ 48.45 & 90.27 $\pm$ 0.015 & 3 & START \\
94.90 $\pm$ 5.85 & 91.12 $\pm$ 0.005 & 4 & START \\ \midrule
99.07 $\pm$ 0.20 & 92.36 $\pm$ 0.017 & 1 & MID \\
99.07 $\pm$ 0.20 & 91.08 $\pm$ 0.011 & 2 & MID \\
99.18 $\pm$ 0.20 & 90.74 $\pm$ 0.012 & 3 & MID \\
40.27 $\pm$ 41.80 & 89.35 $\pm$ 0.026 & 4 & MID \\ \midrule
18.17 $\pm$ 5.08 & 89.00 $\pm$ 0.017 & 1 & END \\
18.28 $\pm$ 7.02 & 89.12 $\pm$ 0.019 & 2 & END \\
75.34 $\pm$ 34.71 & 90.85 $\pm$ 0.008 & 3 & END \\
80.32 $\pm$ 31.39 & 90.16 $\pm$ 0.017 & 4 & END \\ \bottomrule
\end{tabular}
\end{table}

\subsection{Flashy Backdoor}

We merge both concepts presented in the previous attacks, the framing and the strobing, but this time, we fix the trigger size to 100\% of the original frames and ensure there is at least a clean frame between two poisoned frames. This approach ensures that the DVS camera has time to reset its threshold (see Figure~\ref{fig:PoisonframeDVS}) while making the triggers reproducible in real-world scenarios.

As shown in Table~\ref{tab:1} and Table~\ref{tab:3}, our results suggest that combining the full-frame size trigger with the benefits of strobing results in better ASR in every scenario while maintaining the same or even achieving superior clean accuracy in all cases. These results suggest that it is better to concentrate the trigger in fewer frames rather than spreading it through the sample. In Table~\ref{tab:1}, poisoning 30\% of the first 4 frames of the samples obtains substantially better results than poisoning 20\% of 6 frames, even if the poisoning relative to the total amount of pixels in the samples is the same. Additionally, Table~\ref{tab:2} and Table~\ref{tab:4} show that separating the frames in which the trigger is inserted can make the insertion of the backdoor more stable and perform better when inserted in later temporal locations, making it easier to deploy in physical scenarios. Thus, we conclude that apart from being a necessary step for the transference to real-world environments, introducing a full-frame strobing flash is also beneficial in digital scenarios for both the attack's performance and stability.

\subsection{Physical Environment Evaluation}

Lastly, we transfer the \textit{Flashy Backdoor} attack to a physical environment. As our DVS camera had different specifications compared to the one used to capture the DVS128-Gesture dataset, we first started by ensuring our model was able to identify our newly captured samples correctly. We capture both clean and poisoned samples with our DVS camera; the captured samples consist of 4 different subjects, with a total of 89 clean samples and 78 poisoned samples. Using these samples, we created two distinct test sets: one clean and one poisoned, each representing approximately 10\% of the size of the training set used for the models.

Then, we test the initial accuracy of a clean network to validate subsequent
results on the backdoors effectiveness (see Table~\ref{tab:real-world}), obtaining an accuracy of 76.40\% in our clean samples while correctly classifying 61.53\% of the poisoned samples and mistakenly classifying 8.97\% as our trigger label. This behavior is referred to as random guessing, as the model may occasionally predict the trigger label by mistake when classifying other samples. 


\begin{table}
\centering
\caption{Results on real-world samples of the digitally trained models, including a baseline of a clean model measuring the clean accuracy on both clean and poisoned samples (marked with *) and the base ASR.}
\begin{tabular}{@{}ccccc@{}}
\toprule
\textbf{Polarity} & \textbf{Trigger\_length} & \multicolumn{2}{c}{\textbf{Clean acc\%}} & \textbf{ASR\%} \\ \midrule
Baseline          & Baseline                 & 76.40    & 61.53* & 8.97           \\ \midrule
1                 & 1                        & \multicolumn{2}{c}{73.03 $\pm$ 3.37}         & 38.45 $\pm$ 23.11  \\
2                 & 1                        & \multicolumn{2}{c}{53.17 $\pm$ 13.96}        & 65.38 $\pm$ 8.97   \\
3                 & 1                        & \multicolumn{2}{c}{67.79 $\pm$ 3.37}         & 31.96 $\pm$ 30.07  \\ \midrule
1                 & 3                        & \multicolumn{2}{c}{72.65 $\pm$ 3.43}         & 45.72 $\pm$ 34.88  \\
2                 & 3                        & \multicolumn{2}{c}{60.29 $\pm$ 2.34}         & 60.68 $\pm$ 4.50   \\
3                 & 3                        & \multicolumn{2}{c}{71.53 $\pm$ 2.59}         & 97.86 $\pm$ 3.70   \\ \midrule
1                 & 5                        & \multicolumn{2}{c}{70.03 $\pm$ 2.34}         & 43.58 $\pm$ 35.89  \\
2                 & 5                        & \multicolumn{2}{c}{50.18 $\pm$ 11.69}        & 66.23 $\pm$ 8.33   \\
3                 & 5                        & \multicolumn{2}{c}{70.03 $\pm$ 0.64}         & 90.16 $\pm$ 4.85   \\ \bottomrule
\end{tabular}
\label{tab:real-world}
\end{table}

After ensuring that our clean model can correctly identify our real-world samples, we train different models based on the results obtained in the digital scenarios, with polarities $p\in[1,2,3]$. We do not consider $p=0$ as it is not possible to replicate and deploy with a light source; a $p=0$ in all the pixels of the frame would mean that no change in luminosity has occurred from the previous instant.
Our resulting models are trained with poisoned samples of trigger length between 1
and 5 inserted at the start and a gap of size of $1$ between each poisoned frame.

We test each model and observe that the best results are obtained when the trigger polarity is set at $p=3$, meaning that the triggers contained both ON and OFF events, achieving nearly 100\% ASR with little variance when the digital trigger length adapts best to out real triggers (see Table~\ref{tab:real-world}). This result was expected, as most of the poisoned samples captured in the real world contained both ON and OFF events ($p=3$). However, we introduced backdoor behavior across all polarities, even when the trigger did not perfectly match the digital one. 

After analyzing the results, we noticed that increasing the trigger length can result in higher ASR. However, it is necessary to carefully determine the best length, as slightly increasing it can positively affect clean accuracy. However, increasing it too much can also result in a degradation of the clean accuracy depending on the settings. We also notice that, in general, $p=1$ yields the best results in clean accuracy but the worst in ASR. This is due to the triggers having a mix of ON and OFF activations in the majority of the frames, and thus, the model is not able to correctly discern the trigger.
Nevertheless, it is possible to effectively introduce the backdoor behavior while mostly maintaining the initial clean accuracy regardless of the settings, achieving very high clean accuracy and ASR when carefully mimicking the triggers. Thus, we conclude that it is feasible to train a backdoored model with digital triggers and attack the model in real-world scenarios without the need to introduce any physical trigger sample in the training data or perfectly match the digital trigger. 

\subsection{Evaluation Against Backdoor Defenses}

Given the lack of specific backdoor defenses for SNNs, we experiment with two commonly used methods in DL that have been studied as a means to enhance the security of already trained models, i.e., pruning and fine-tuning, and apply a defense from the DNN domain to work with neuromorphic data in SNNs: fine-pruning~\cite{fine_prune}.

\begin{table*}[t]
\centering
\caption{Results of the defense procedures for all the presented attacks and the DVS128-Gesture, NMnist, and Cifar10-DVS datasets.}\label{table11}
\renewcommand{\arraystretch}{1.2}
\resizebox{\textwidth}{!}{
\fontsize{14pt}{14pt}\selectfont
\begin{tabular}{@{}cccccccccccccc@{}}
\toprule
\textbf{Dataset} & \textbf{Size} & \textbf{Attack} & \makecell{\textbf{Poisoned} \\ \textbf{Frames}} & \textbf{Ini CA\%} & \textbf{I ASR\%} & \textbf{$\tau$} & \textbf{Pruned CA\%} & \textbf{Pruned ASR\%} & \makecell{\textbf{FT} \\ \textbf{epochs}} & \textbf{FT CA\%} & \textbf{FT ASR\%} & \textbf{FP CA\%} & \textbf{FP ASR\%} \\ \midrule
DVS128-Gesture & 0.3 & Static & 16 & 90.62 $\pm$ 0.35 & 99.08 $\pm$ 0.40        & 0.8 & 24.31 $\pm$ 13.02 & 9.38 $\pm$ 13.84  & 4 & 88.08 $\pm$ 2.12 & 68.98  $\pm$ 7.29  & 86.34 $\pm$ 1.75 & 17.71 $\pm$ 3.63   \\
DVS128-Gesture & 0.3 & Framed & 12 & 91.20 $\pm$ 1.00 & 99.08 $\pm$ 0.20        & 0.8 & 31.60 $\pm$ 3.67 & 8.94 $\pm$ 11.09   & 4 & 89.70 $\pm$ 1.64 & 70.25  $\pm$ 5.95  & 90.74 $\pm$ 2.44 & 21.99 $\pm$ 1.56   \\
DVS128-Gesture & 0.3 & Framed & 6 & 89.47 $\pm$ 1.56 & 54.63 $\pm$ 35.50        & 0.8 & 41.90 $\pm$ 5.88 & 9.49 $\pm$ 14.94   & 4 & 85.88 $\pm$ 5.23 & 14.59  $\pm$ 3.31  & 87.96 $\pm$ 3.11 & 10.19 $\pm$ 2.26   \\
DVS128-Gesture & 0.3 & Strobing & 6 & 89.00 $\pm$ 1.56 & 97.80 $\pm$ 0.87       & 0.8 & 22.55 $\pm$ 3.75 & 0.70 $\pm$ 0.92    & 4 & 87.15 $\pm$ 3.62 & 21.30  $\pm$ 16.47 & 89.47 $\pm$ 1.71 & 11.11 $\pm$ 4.52   \\
DVS128-Gesture & 1 & Flash & 6 & 90.20 $\pm$ 0.92 & 99.54 $\pm$ 0.53            & 0.8 & 48.38 $\pm$ 12.14 & 99.54 $\pm$ 0.80  & 4 & 86.34 $\pm$ 2.63 & 47.22  $\pm$ 19.95 & 86.80 $\pm$ 3.31 & 14.82 $\pm$ 4.18   \\
DVS128-Gesture & 1 & Flash & 3 & 87.61 $\pm$ 1.91 & 96.64 $\pm$ 4.07            & 0.8 & 36.81 $\pm$ 15.14 & 69.91 $\pm$ 40.36 & 4 & 87.62 $\pm$ 0.87 & 15.63  $\pm$ 5.46  & 81.25 $\pm$ 1.59 & 8.68 $\pm$ 1.39    \\
DVS128-Gesture & 1 & Str. Flash & 3 & 88.31 $\pm$ 2.81 & 90.28 $\pm$ 15.04  & 0.8 & 35.53 $\pm$ 7.46 & 55.90 $\pm$ 42.02  & 4 & 89.46 $\pm$ 1.22 &  23.61 $\pm$ 12.28 & 87.50 $\pm$ 2.84 & 8.45 $\pm$ 1.11    \\ \midrule
NMnist & 0.3 & Static & 16 & 99.28 $\pm$ 0.03 & 99.97 $\pm$ 0.00                & 0.8 & 88.60 $\pm$ 9.19 & 99.35 $\pm$ 0.97   & 2 & 99.05 $\pm$ 0.12 &  99.93 $\pm$ 0.07  & 98.81 $\pm$ 0.12 & 32.99 $\pm$ 15.03  \\
NMnist & 0.3 & Framed & 10 & 99.28 $\pm$ 0.03 & 99.97 $\pm$ 0.00                & 0.8 & 85.20 $\pm$ 7.74 & 99.06 $\pm$ 0.60   & 2 & 99.08 $\pm$ 0.11 &  99.97 $\pm$ 0.01  & 98.72 $\pm$ 0.25 & 28.42 $\pm$ 15.73  \\
NMnist & 0.3 & Framed & 5 & 99.24 $\pm$ 0.03 & 99.97 $\pm$ 0.01                 & 0.8 & 92.28 $\pm$ 2.80 & 81.37 $\pm$ 28.70  & 2 & 99.19 $\pm$ 0.03 &  99.97 $\pm$ 0.07  & 98.80 $\pm$ 0.13 & 21.84 $\pm$ 2.56   \\
NMnist & 0.3 & Strobing & 5 & 99.29 $\pm$ 0.04 & 99.97 $\pm$ 0.01               & 0.8 & 79.21 $\pm$ 6.95 & 98.33 $\pm$ 2.67   & 2 & 99.01 $\pm$ 0.15 &  99.83 $\pm$ 0.25  & 98.83 $\pm$ 0.08 & 49.01 $\pm$ 23.06  \\
NMnist & 1 & Flash & 6 & 99.29 $\pm$ 0.03 & 99.97 $\pm$ 0.02                    & 0.8 & 89.01 $\pm$ 4.40 & 69.82 $\pm$ 52.05  & 2 & 99.05 $\pm$ 0.07 &  99.96 $\pm$ 0.01  & 98.74 $\pm$ 0.16 & 99.95 $\pm$ 0.04   \\
NMnist & 1 & Flash & 3 & 99.27 $\pm$ 0.02 & 99.97 $\pm$ 0.00                    & 0.8 & 94.40 $\pm$ 2.09 & 78.06 $\pm$ 25.44  & 2 & 99.01 $\pm$ 0.08 &  99.97 $\pm$ 0.01  & 98.87 $\pm$ 0.06 & 83.45 $\pm$ 20.19  \\
NMnist & 1 & Str. Flash & 3 & 99.30 $\pm$ 0.02 & 99.96 $\pm$ 0.01           & 0.8 & 87.26 $\pm$ 9.32 & 69.35 $\pm$ 48.22  & 2 & 99.02 $\pm$ 0.23 &  99.95 $\pm$ 0.06  & 98.82 $\pm$ 0.13 & 60.95 $\pm$ 40.13  \\ \midrule
Cifar10-DVS & 0.3 & Static & 16 & 67.47 $\pm$ 0.78 & 95.83 $\pm$ 0.06           & 0.8 & 33.27 $\pm$ 1.63 & 70.80 $\pm$ 16.75  & 2 & 65.93 $\pm$ 0.95 &  85.30 $\pm$ 2.43  & 65.20 $\pm$ 1.51 & 29.43 $\pm$ 13.31  \\
Cifar10-DVS & 0.3 & Framed & 10 & 67.30 $\pm$ 0.90 & 95.67 $\pm$ 0.06           & 0.8 & 27.53 $\pm$ 0.74 & 92.67 $\pm$ 2.25   & 2 & 66.13 $\pm$ 0.85 &  90.87 $\pm$ 6.20  & 65.23 $\pm$ 1.52 & 56.57 $\pm$ 12.68  \\
Cifar10-DVS & 0.3 & Framed & 5 & 66.33 $\pm$ 0.61 & 95.60 $\pm$ 0.10            & 0.8 & 28.07 $\pm$ 2.11 & 94.40 $\pm$ 0.92   & 2 & 64.37 $\pm$ 1.02 &  87.63 $\pm$ 6.55  & 62.70 $\pm$ 3.05 & 64.83 $\pm$ 5.18   \\
Cifar10-DVS & 0.3 & Strobing & 5 & 66.20 $\pm$ 0.72 & 95.57 $\pm$ 0.21          & 0.8 & 30.23 $\pm$ 2.25 & 94.80 $\pm$ 2.30   & 2 & 64.30 $\pm$ 1.54 &  95.23 $\pm$ 2.51  & 63.67 $\pm$ 1.16 & 95.40 $\pm$ 0.89   \\
Cifar10-DVS & 1 & Flash & 6 & 63.40 $\pm$ 0.98 & 95.87 $\pm$ 0.15               & 0.8 & 11.07 $\pm$ 1.22 & 99.90 $\pm$ 0.17   & 2 & 61.67 $\pm$ 1.47 &  95.40 $\pm$ 0.17  & 63.03 $\pm$ 1.16 & 95.40 $\pm$ 0.89   \\
Cifar10-DVS & 1 & Flash & 3 & 63.97 $\pm$ 0.93 & 95.57 $\pm$ 0.42               & 0.8 & 17.20 $\pm$ 9.68 & 98.50 $\pm$ 2.18   & 2 & 62.77 $\pm$ 1.00 &  85.90 $\pm$ 9.01  & 61.60 $\pm$ 0.17 & 55.73 $\pm$ 10.23  \\
Cifar10-DVS & 1 & Str. Flash & 3 & 64.47 $\pm$ 1.12 & 95.53 $\pm$ 0.21      & 0.8 & 12.30 $\pm$ 3.64 & 99.87 $\pm$ 0.23   & 2 & 61.63 $\pm$ 1.80 &  95.80 $\pm$ 0.26  & 63.70 $\pm$ 0.85 & 92.30 $\pm$ 4.16   \\\bottomrule
\end{tabular}
}
\end{table*}

\subsubsection{Pruning}

Pruning is commonly used to reduce the computational expense of DNNs~\cite{pruning_1, pruning_2, pruning_3}, showing that a significant amount of neurons can be pruned before starting to compromise the classification accuracy. However, it was also explored as a method to enhance the security of corrupt DNNs by Liu et al.~\cite{fine_prune}.
The premise behind this defense is that the attacked model learns to misbehave on backdoored inputs while behaving correctly on clean inputs because there are ``backdoor neurons'' that remain dormant in the presence of clean inputs but activate under the presence of the specified trigger, which was empirically demonstrated by Gu et al.~\cite{BadNets}. 


We apply this procedure to the last convolutional layer of the model with a pruning rate of $\tau = 0.8$, meaning that 80\% of the neurons in that layer are pruned. 
We also focus on the effect that the length of the trigger and the clean gap has in each proposed attack on the DVS128-Gesture, NMnist, and Cifar10-DVS datasets (see Table~\ref{table11}). As can be seen, even though the initial clean accuracy and ASR are high before pruning, we observe a substantial decrease in the clean accuracy across all datasets (except on NMnist, where the drop is less noticeable) after pruning.

Upon examining these results,
a general tendency can be observed, as pruned models trained with bigger triggers tend to have a more substantial drop in clean accuracy and less impact on ASR. This outcome was expected; a smaller trigger can be more discrete, but it also has less impact on the model, mainly injecting the backdoor on less influential neurons. The pruning process first focuses on removing these less influential neurons, making it easier to erase the backdoor behavior when the triggers are small. However, with the introduction of larger triggers, their influence on the model is also amplified, making them more challenging to target through this procedure.

\subsubsection{Fine-Tuning}
\label{tab:prune_defense}
In general, fine-tuning refers to taking a pre-trained neural network model and further training it on a new dataset or task. The pre-trained model is typically initialized with weights learned from training on a large, general dataset. During fine-tuning, the parameters of the pre-trained model are adjusted or fine-tuned to adapt to the specifics of the new dataset or task. This adjustment allows the model to learn task-specific features and patterns while leveraging the knowledge captured by the pre-trained model. 

In our case, we take the given backdoored model and retrain it for a fixed amount of epochs on a clean dataset, allowing the model to try to ``forget'' the backdoor behavior. This is based on the concept of ``catastrophic forgetting'', which refers to the phenomenon where a neural network model trained on a specific task or dataset forgets previously learned information when trained on a new task or dataset~\cite{forgetting}. We fine-tune the model for 10\% of the network's original training epochs with the clean DVS128-Gesture, NMnist, and Cifar10-DVS datasets, focusing again on each attack type's trigger length and size (see Table~\ref{table11}). This decision was taken based on both prior studies that showed a significant ASR reduction with a small amount of fine-tuning epochs~\cite{few_epochs1, fine_prune} and the threat model, as the user who requested the service would not have the resources needed for extensive training. 



Our results suggest that fine-tuning the models for a small number of epochs can reduce the backdoor effect for certain datasets and attack configurations, particularly those involving smaller triggers and shorter trigger lengths on the DVS128-Gesture dataset. However, longer triggers and larger trigger sizes tend to be more resilient, showing less reduction in ASR after fine-tuning. It is also noteworthy that including a clean frame between triggers resulted in higher ASRs across all tested datasets than their continuous counterparts. 
\subsubsection{Fine-Pruning}
This defense~\cite{fine_prune} is a combination of both pruning and fine-tuning. The defender first prunes the network and then fine-tunes the result. The main intuition behind this defense is that the initial pruning removes the backdoor neurons, and the following fine-tuning restores the drop in classification accuracy on the clean inputs introduced by the pruning. 

Even if this defense can be effective when the trigger size is small, as demonstrated by Abad et al.~\cite{paperGorka2}, when applied against larger triggers, the results are not straightforward. We can observe a higher success when compared to the previous two methods, achieving a substantial reduction in ASR and being able to recover almost all of the clean accuracy in most cases. But when faced with bigger triggers in NMist or Cifar10-DVS datasets, it is not able to completely eliminate the backdoor behavior (see Table~\ref{table11}). 
We hypothesize that the reason this defense performs better than fine-tuning alone, even though pruning by itself does not seem to be effective, is that pruning the less important neurons allows the fine-tuning process to focus on the more critical neurons where the backdoor is embedded. By eliminating the neurons that contribute least to the model's overall performance, pruning reduces the model's capacity to retain the backdoor behavior. This targeted approach enables the fine-tuning phase to overwrite the backdoor-influenced parameters, thus enhancing the defense's effectiveness.



It is also to be noted that fine-pruning focuses on identifying and nullifying the model parameters that introduce the malicious behavior. However, as demonstrated in previous works~\cite{paperGorka2}, when faced with an adaptive attacker, it becomes easily evaded. As discussed by Abad et al.~\cite{paperGorka2}, an attacker who knows that the last convolutional layer is targeted for pruning could limit the backdoor injection to the rest of the network or reduce the poisoning rate, achieving high ASR and nullify the pruning procedure.

\subsection{Evaluation of the Stealthiness}

Assessing the stealthiness of triggers in backdoor attacks is critical to ensure their effectiveness while remaining undetected. In practical applications, the datasets employed are often extensive~\cite{VeryBigData}, making it impractical to visually evaluate the presence of triggers by human users. Manual inspection would be not only time-consuming~\cite{BigDataTimeConsuming} but also prone to human error and subjectivity~\cite{humanError}, making it an inefficient method for large-scale datasets~\cite{LargeScale}.

To address this challenge, quantitative metrics are utilized to evaluate the imperceptibility of the backdoor triggers objectively. These metrics provide a systematic approach to measure the similarity between the original, unaltered images and the images containing the backdoor triggers, ensuring that the triggers are not easily detectable by automated systems. Some of the commonly used metrics are the Mean Squared Error (MSE), the Structural Similarity Index Measure (SSIM), and the Peak Signal-to-Noise Ratio (PSNR)~\cite{common_metrics}. We additionally explore other methods, such as an entropy evaluation of the activation values, the number of activations across a whole frame, and the entropy of those activations among each frame.

\subsubsection{Mean Squared Error}
The MSE~\cite{mse} is a widely used metric that calculates the average square values of the differences between the corresponding pixel values of the original and the modified images.
A lower MSE value indicates higher similarity between the images, suggesting that the trigger is less perceptible. However, this metric is discussed not to be a good representation of how a human user would discern the change in the image~\cite{mse}, as it does not take into account the context of the neighboring pixels. This metric would give the same result for a change of 10\% of the total pixel amount distributed across the image or all of them focused on the same spot, resulting in an evident change for a human user but resulting in the same MSE.

\subsubsection{Structural Similarity Index Measure}

SSIM, proposed by Wang et al.~\cite{ssim}, is another widely used metric that assesses the visual impact of changes in structural information between two images~\cite{common_metrics}. Unlike MSE, SSIM considers luminance, contrast, and structure changes, making it more aligned with human visual perception. The SSIM index ranges from -1 to 1, where 1 indicates perfect structural similarity. 

\subsubsection{Peak Signal-to-Noise Ratio}

PSNR is a commonly used metric to evaluate the quality of an image by comparing the original and the altered image~\cite{psnr}. Opposed to the MSE metric, it takes into account images having different dynamic ranges~\cite{mse}, which in this case can be very beneficial since the pixels have no set range; it depends on the number of activations that occurred on the selected time frame. This metric expresses the ratio between the maximum possible power of a signal and the power of corrupting noise that affects the fidelity of its representation. PSNR is expressed in decibels (dB) and is inversely related to the MSE. 

PSNR is particularly useful in scenarios where the human perception of image quality is crucial. It is extensively used in image compression, transmission, and processing to evaluate how much an image has been degraded by these processes~\cite{psnr}. In the context of backdoor attacks, a high PSNR value would suggest that the trigger is well-hidden and that it does not significantly degrade the image quality, thus remaining stealthy.

\subsubsection{Entropy of the Activation Values}

Entropy measures the amount of randomness in the pixel values of an image. Lower entropy values in the trigger image, when compared to the base image, would suggest that the trigger introduces substantial changes as it does not maintain the overall statistical properties of the original image. 

\begin{figure}[ht]
    \centering
    \includegraphics[width = 1\linewidth, trim= 28pt 29pt 0pt 35pt, clip]{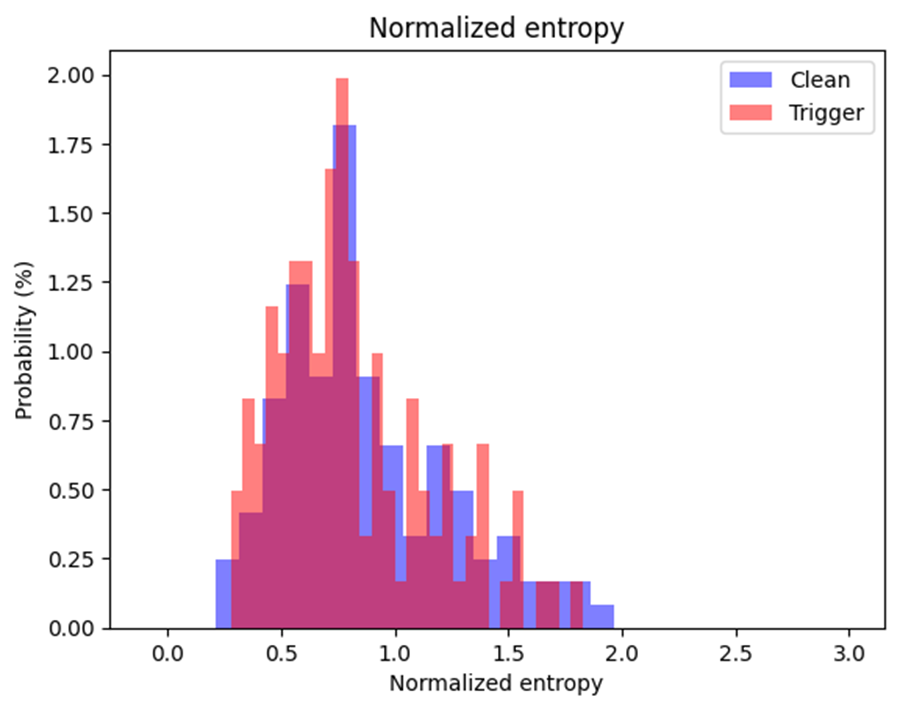}
    \caption{Normalized entropy of the pixel values across all samples. Represented in blue are the clean sample pixels' values, and in red are the trigger ones. The X-axis depicts the pixel values while the Y-axis the probability of appearance.}
    \label{fig:NormEntropy}
\end{figure}

\subsubsection{Number of Activations per Channel}
We explore the option of examining the average number of activations per channel in each frame of the images. The intuition behind this analysis is that including a large trigger in the image would significantly change the number of activations in the frames where it is present. 
\begin{figure}[ht]
    \centering

    
    \begin{subfigure}[b]{0.49\linewidth}
        \includegraphics[width=\linewidth, trim= 0pt 5pt 5pt  0pt, clip]{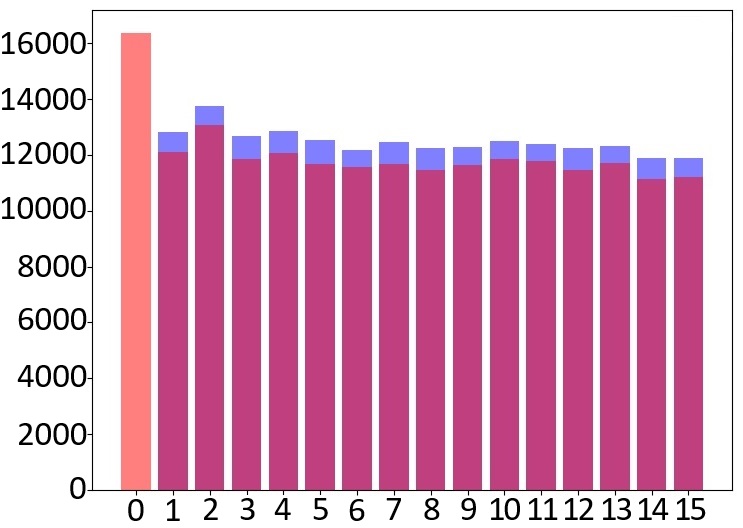}
        \caption{Poison p=1}
    \end{subfigure}
    \hfill
    \begin{subfigure}[b]{0.49\linewidth}
        \includegraphics[width=\linewidth]{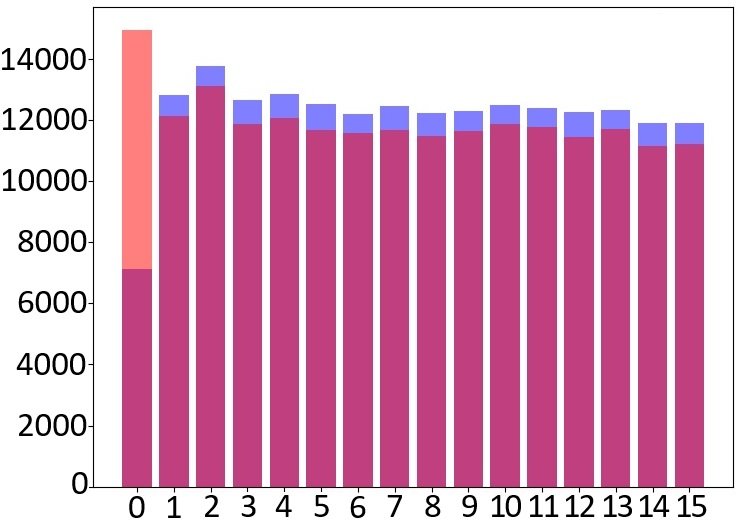}
        \caption{Mixed p=1}
    \end{subfigure}

    \begin{subfigure}[b]{0.49\linewidth}
        \includegraphics[width=\linewidth]{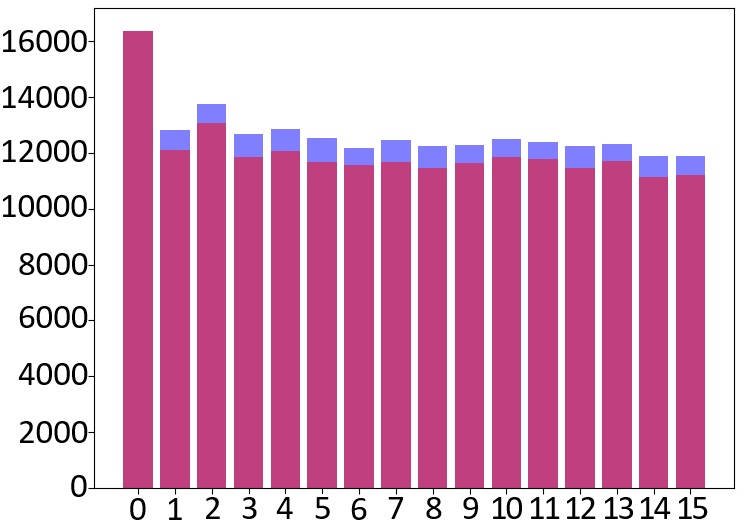}
        \caption{Poison p=3}
    \end{subfigure}
    \hfill
    \begin{subfigure}[b]{0.49\linewidth}
        \includegraphics[width=\linewidth, trim= 35pt 10pt 5pt  10pt, clip]{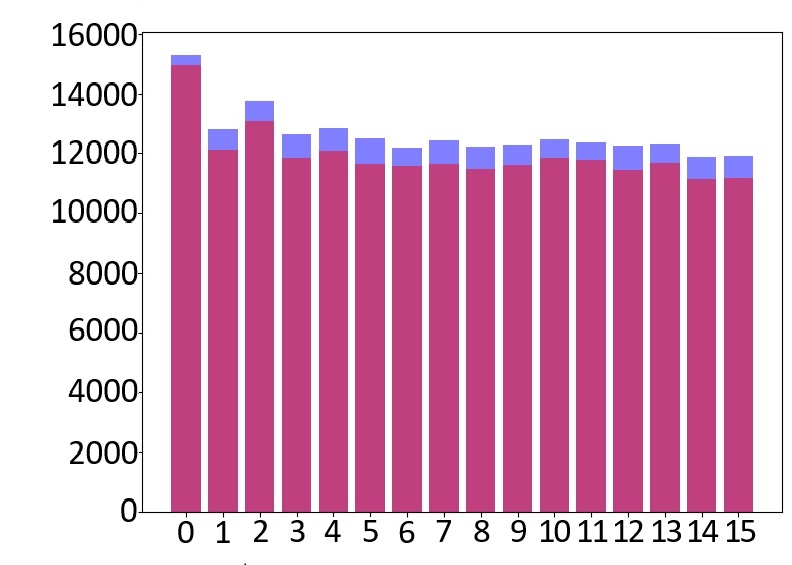}
        \caption{Mixed P=3}
    \end{subfigure}
    \caption{Mean number of activations per channel and frame on poisoned, mixed, and clean samples where the trigger is inserted in the first frame with polarities 1 and 3. In all graphs, the X-axis represents the frame number, and the Y-axis denotes the number of activations. OFF activations are depicted in red, while ON activations are shown in blue.}
    \label{fig:meanAct}
\end{figure}

Additionally, we also examine the entropy of the average activation per frame. The concept is that frames containing triggers would exhibit similar activation numbers between them on the poisoned frames when compared to the clean frames.

\begin{figure}[ht]
    \centering
    \begin{subfigure}[b]{0.49\linewidth}
        \centering
        \includegraphics[width=\linewidth]{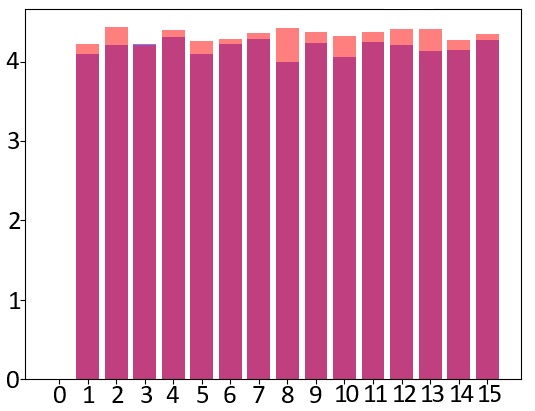}
        \caption{Poisoned}
    \end{subfigure}
    \hfill
    \begin{subfigure}[b]{0.49\linewidth}
        \centering
        \includegraphics[width=\linewidth]{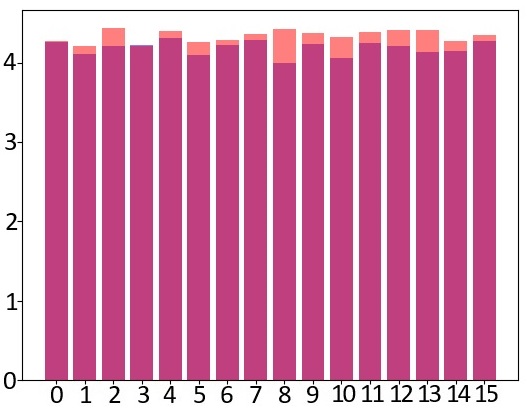}
        \caption{Mixed}
        
    \end{subfigure}
    \caption{Entropy of the number of activations in each frame of poisoned, mixed, and clean samples for triggers with $p=3$. The X-axis represents the frame number, and the Y-axis denotes the entropy of the activations. OFF activations are depicted in red, while ON activations are shown in blue.}
    \label{fig:actEntropy}
\end{figure}

\subsubsection{Evaluation Results}
We evaluate these metrics in a test set consisting of 116 clean images and the same amount of poisoned ones. The poisoned samples are generated by inserting a full frame trigger on the first frame of every sample in the clean set with $p=3$.

\begin{table}[ht]
    \caption{Evaluation of the stealthiness metrics on the clean and poisoned dataset.}
    \centering
    \begin{tabular}{ccc}
    \hline
    \multicolumn{2}{c}{\textbf{Metric}} & \textbf{Value} \\ \hline
    \multicolumn{1}{c|}{\multirow{2}{*}{Mean Pixel Entropy}} & Clean & 0.87 \\
    \multicolumn{1}{c|}{} & Trigger & 0.81 \\
    Average MSE &  & 0.44 \\
    PSNR &  & 40.06 \\
    SSIM &  & 0.95 \\ \hline
    \end{tabular}
    
    \label{tab:stealth}
\end{table}

As seen in Table~\ref{tab:stealth}, there is no obvious way to determine if the dataset contains poisoned samples. There is little difference in the pixel entropy between the clean and the poisoned samples (see Figure~\ref{fig:NormEntropy}). The average MSE could indeed be considered relatively high, but the PSNR, which is inversely related to the MSE, obtained a mark of 40.06; in this metric, anything above 40db is considered to be very good as it means that the reconstruction of the image contains little noise added. This difference in these two metrics is due to the nature of neuromorphic data, as its pixel values have no upper bound; each pixel contains the number of activations in a set amount of time. SSIM, on the other hand, obtained a value of 0.95, which indicates a nearly perfect structural similarity.

After seeing that these metrics were ineffective in discerning the poisoned samples, we focused on analyzing the number of activations per frame instead of the whole sample. Depending on the polarity and the size of the trigger, this could be a method of identifying a change in specific frames of some samples. However, when the polarity is set to $p=3$, which also obtained the best results on ASR in real-world environments (see Table~\ref{tab:real-world}), it becomes much more challenging to discern the insertion of the triggers when the poisoned samples are mixed with the clean ones (see Figure~\ref{fig:meanAct}). It is also to be noted that this comparison was made using a ratio of 1:1 between clean and poisoned samples, when in reality, the dataset would contain only one poisoned sample for every ten clean ones, making it even harder to discern. 

When analyzing the entropy of each frame on a single sample by itself, we found that it is easy to discern the triggered frames when the size of the trigger represents the majority of the image, as the entropy of that frame would be very low or 0 if the trigger only contains a single value across the whole frame. However, this means that the user would need to analyze each frame of each sample by itself, which would be highly unpractical for large datasets or samples with more significant numbers of frames. If we apply this method to the whole dataset at once, we find no observable difference when comparing clean frames to the triggered ones (see Figure~\ref{fig:actEntropy}).


\section{Conclusions \& Future Work}
Even though the vulnerabilities
of traditional DNNs have been extensively studied, the susceptibility of SNNs to backdoor attacks remains mostly underexplored, especially in physical scenarios,
where no studies had been conducted. This paper addresses this gap by investigating the potential for an attacker to digitally craft and inject a backdoor in an SNN model to exploit it in real-world scenarios. We introduce and evaluate three different backdoor attack methodologies, i.e., \emph{Framed Backdoor}, \emph{Strobing Backdoor}, and \emph{Flashy Backdoor}. Each is designed to exploit or evade unique aspects of SNNs and DVS cameras, making the final attack applicable to physical environments.

The first methodology, \emph{Framed Backdoor}, which focuses on static triggers positioned at different intervals within samples, demonstrates a high success rate in achieving 100\% ASR and minimal clean accuracy degradation on the attacked model with a small number of frames required for the trigger. It also enhances the overall stealthiness of the triggers, requiring only a fraction of the total sample frames to be poisoned for the attack to achieve top performance. However, it is not applicable in physical environments as it cannot evade the adaptive nature of the DVS cameras.

The second one, \emph{Strobing Backdoor}, introduces non-continuous triggers, which enables the attacker to evade the adaptative threshold of DVS cameras and make the triggers reliably reproducible in physical scenarios. This approach demonstrated better capabilities in the majority of scenarios, requiring a smaller amount of frames to achieve comparable ASR to its continuous counterpart while also maintaining the same clean accuracy. This methodology solves the aforementioned problem; however, the generated triggers are still not easily and reliably reproducible in real-world scenarios; thus, we present our last attack methodology, the \emph{Flashy Backdoor}.

The \emph{Flashy Backdoor} combines both concepts to generate the first real-world environment backdoor attack on SNNs, enabling the attacker to digitally craft and introduce the triggers in the model and exploit the introduced backdoor in the real world without the need to insert any real poisoned samples. This method showed excellent performance both in digital and physical environments, achieving a 100\% ASR while retaining most of the clean accuracy.

Additionally, we evaluate the stealthiness of the triggers using common metrics such as MSE, SSIM, and PSNR. Moreover, we propose alternatives more suited for these attacks on neuromorphic data, like the analysis of the number of activations per channel or the entropy distribution of the activation values per frame. We found that these traditional metrics are not effective in discerning the poisoned samples from the real ones. And, while with our proposed methods, the poisoned samples can be identified when isolated, stronger methods need to be found to be able to identify them when mixed with clean samples (as they would commonly be when inserted on a dataset).

This research also assesses the effectiveness of traditional DNN procedures and backdoor defenses on SNNs, showing limited effectivity and often sacrificing too much clean accuracy to be able to be actually employed, emphasizing the need for stronger SNN-focused defenses. 
Our findings prove the vulnerabilities and the ability of an attacker to digitally inject a backdoor on a real-world-oriented SNN system without the need to capture any physical triggers for it to be effective when deployed in a real-world environment.

Future research should focus on finding more advanced detection mechanisms, as well as more robust defense strategies targeted for SNNs. Additionally, the proposed attacks should also be tested across a broader range of scenarios and datasets to validate their generalizability and robustness. This includes exploring varying environmental conditions that could be present in real-world scenarios, like changing lighting conditions, background activity, or adverse meteorological conditions.

\bibliographystyle{plain}
\bibliography{Bibliography}

\appendices

\section{Data Availability} 
\label{dataAvailability}
The dataset and code used in this study are available in the following anonymous repository. This includes the code, experimental setups, and data samples used for the evaluation.
Due to the large size of the real-world captured samples, they have been hosted on a separate cloud storage service. Interested researchers can access this data via the following~\href{https://www.dropbox.com/scl/fi/sc2640xnyg6ojo0bjlyju/Physical_sample_dataset.tar?rlkey=ddr7dnhug67e5dbhmii9nk6df&st=8x4e2c84&dl=0}{link}\footnote{\url{https://www.dropbox.com/scl/fi/sc2640xnyg6ojo0bjlyju/Physical_sample_dataset.tar?rlkey=ddr7dnhug67e5dbhmii9nk6df&st=8x4e2c84&dl=0}}.
The shared data includes all materials necessary to reproduce our experiments, including poisoned samples and real-world backdoor attack implementations. The use of the dataset is intended for academic research only.


\section{Dataset and Model Characterization}
\label{appendixB}
Here, we discuss the details of the datasets, the architectural details of each model, and the training settings used on each one. We use three different architectures employed in related works~\cite{enhance_SNN}, one for each dataset. 

\subsection{N-MNIST}
The N-MNIST dataset, introduced by Orchard et al.~\cite{N-MNIST}, is a neuromorphic version of the traditional MNIST dataset designed to be used with SNNs and neuromorphic vision systems. However, this dataset retains the original $28\times28$ pixel resolution of the MNIST dataset, presenting it in a spiking format suitable for neuromorphic processing.
Each event in the N-MNIST dataset is represented by a 40-bit word that includes the $X$ and $Y$ addresses (each 8 bits), the polarity (1 bit, indicating whether the event is an ON or OFF event), and a timestamp (23 bits, in microseconds). On average, each N-MNIST recording contains approximately 4,172 events, with an almost equal number of ON and OFF events. These events provide spatial and temporal information about the pixel changes, which are crucial for the performance of SNNs.
The dataset is organized into directories corresponding to each digit class (0-9), and it includes separate training and testing sets, mirroring the original MNIST structure. Each image is stored as a binary file containing the list of events, preserving the temporal dynamics necessary for neuromorphic vision.

\subsubsection{Architecture And Parameters}
\begin{compactitem}
    \item Input Channels: The network begins with a single-channel input, typical for grayscale images like those in the MNIST dataset.
    \item Convolutional Layers: There are two main convolutional blocks in this network:
    \begin{compactitem}
        \item First Block:
        \begin{compactitem}
            \item A 2D convolutional layer (Conv2d) with 2 input channels and 128 output channels, using a kernel size of $3\times3$, padding of 1, and no bias.
            \item A batch normalization layer (BatchNorm2d) to stabilize the learning process by normalizing the activations.
            \item A spiking neuron layer processes the output of the batch normalization layer, adding non-linearity and temporal dynamics.
            \item A max pooling layer (MaxPool2d) with a kernel size of $2\times2$ to reduce the spatial dimensions of the output.
            \end{compactitem}
        \item Second Block:
        \begin{compactitem}
            \item Another Conv2d layer with both input and output channels set at 128, also using a kernel size of 3x3, padding of 1, and no bias.
            \item This is again followed by a BatchNorm2d and another spiking neuron layer.
            \item Another max pooling layer to reduce the dimensions.
            \end{compactitem}
    \end{compactitem}
    \item Flatten and Dropout Layers: The output from the convolutional layers is flattened into a vector and processed through a dropout layer with a rate of 0.5 to prevent overfitting.
    \item Fully Connected Layers:
    \begin{compactitem}
        \item The first dense layer (Linear) maps the flattened vector to 2048 neurons.
        \item After another spiking neuron layer and a dropout layer, the second dense layer maps these 2048 neurons down to 100.
        \end{compactitem}
    \item Voting Layer: The final layer outputs the class predictions. It maps the 100 output features to 10 classes corresponding to the digits 0 through 9.
\end{compactitem}
All models trained with this architecture for the N-MNIST dataset were trained for ten epochs.

\subsection{Cifar10-DVS}
The CIFAR10-DVS dataset, presented by Li et al.~\cite{CIFAR_DVS}, is a neuromorphic version of the CIFAR-10 dataset tailored for event-based vision research. This dataset consists of 10,000 event streams converted from the original CIFAR-10 images using a DVS.
Each event in the CIFAR10-DVS dataset is encoded with the $X$ and $Y$ addresses (each 7 bits), the polarity (1 bit, indicating ON or OFF events), and a timestamp (24 bits, in microseconds). The dataset is structured into 10 classes, with 1,000 images per class, randomly selected from the original CIFAR-10 dataset. The conversion process involves upscaling the original $32\times32$ color images to $512\times512$ pixels using bicubic interpolation to ensure clear recordings when displayed on an LCD monitor. It is stored similarly to the N-MNIST dataset.

\subsubsection{Architecture And Parameters}
\begin{compactitem}
    \item Input Channels: The initial input to the network has 2 channels.
    \item Convolutional Layers: The network includes four blocks, each consisting of:
        \begin{compactitem}
        \item A 2D convolutional layer (Conv2d) with a kernel size of $3\times3$, padding of 1, and no bias. The first layer starts with 2 input channels, while the subsequent layers use 128 channels.
        \item A batch normalization layer (BatchNorm2d) that helps in normalizing the activations of the previous layer, improving stability and performance.
        \item A spiking neuron layer that introduces non-linearity and temporal dynamics into the network.
        \item A 2D max pooling layer (MaxPool2d) with a kernel size of $2\times2$ that reduces the dimensionality of the data for computational efficiency and abstraction.
        \end{compactitem}
    \item Flatten and Dropout Layers: After the convolutional blocks, the network flattens the data into a single vector and applies a dropout (0.5 rate) to reduce overfitting.
    \item Fully Connected Layers: There are two fully connected (Linear) layers:
        \begin{compactitem}
        \item The first maps the flattened feature set to 512 neurons.
        \item The second reduces these to 100 neurons.
        \end{compactitem}
    \item Additional Spiking Neurons: Spiking neurons are again used after each fully connected layer to maintain consistent neural dynamics throughout the network.
    \item Voting Layer: A final voting layer outputs the predictions across 10 classes.
\end{compactitem}
All models trained with this architecture for the Cifar10-DVS dataset were trained for 20 epochs.

\subsection{DVS128-Gesture}
The DVS128-Gesture dataset, introduced by Amir et al.~\cite{gestures}, is designed for gesture recognition using event-based processing. The dataset was recorded using the iniLabs DVS128 camera, which has a resolution of $128\times128$ pixels. The camera setup involved observing gestures performed by subjects standing against a stationary background. Each event includes the $X$ and $Y$ coordinates of the pixel that changed, the polarity of the change (ON or OFF), and a timestamp indicating the exact time of the event.
The dataset comprises 11 hand and arm gesture categories: hand waving (both arms), large straight arm rotations (clockwise and counterclockwise, both arms), forearm rolling (forward and backward), air guitar, air drums, and an “Other” category where subjects could perform a gesture of their choice. Data was collected from 29 subjects under three different lighting conditions (natural light, fluorescent light, and LED light), with each subject performing all 11 gestures sequentially in a single session. This resulted in 1,342 gesture instances grouped into 122 samples.
The events are recorded with microsecond precision, capturing the fine temporal details of the gestures. Each gesture instance is recorded as a series of timestamped events; unlike the other two datasets, this time, the samples of each subject are recorded as a continuous stream, and a CSV details the specific timestamp where each gesture is being performed.

\subsubsection{Architecture And Parameters}
\begin{compactitem}
    \item Input Channels: The network starts with an input with 2 channels, the X and Y dimensions of a frame from a DVS camera.
    \item Convolutional Layers: The network consists of five blocks, each containing: 
    \begin{compactitem}
        \item A 2D convolutional layer (Conv2d) that progressively takes in more channels. The first block starts with 2 input channels, and subsequent blocks use 128 channels. Each convolutional layer uses a kernel size of $3\times3$ with padding of 1 and no bias.
        \item A batch normalization layer (BatchNorm2d) to stabilize and speed up training.
        \item A spiking neuron layer that introduces non-linear, dynamic behavior typical of spiking neural networks.
        \item A 2D max pooling layer (MaxPool2d) with a kernel size of $2\times2$ to reduce the spatial dimensions of the output.
    \end{compactitem}
    \item Flatten and Dense Layers: After the convolutional blocks, the data is flattened to prepare it for fully connected layers.
    \item Dropout: Dropout layers with a rate of 0.5 are added before and after the first fully connected layer to prevent overfitting.
    \item Fully Connected Layers: There are two fully connected (Linear) layers:
     \begin{compactitem}
        \item The first one maps the flattened features to 512 neurons.
        \item The second one maps these 512 neurons to 110 neurons.
    \end{compactitem}
    \item Spiking Neurons: These are also added after each fully connected layer to maintain the dynamic and non-linear properties of the network.
    \item Voting Layer: Finally, a voting layer consolidates the outputs into 10 classes, which correspond to different gestures.
\end{compactitem}
All models trained with this architecture for the DVS128-Gesture dataset were trained for either 40 or 80 epochs, the latter being for real-world environments, as the increase in the number of epochs did not affect the digital environments but positively influenced the base accuracy in real ones.

\section{Parameter Tuning and Biases of the DVS Camera} 
\label{appendixC}
\label{appendixD}
This section collects all the biases and parameters used to capture the samples in our specific environment, as well as a brief explanation of the influence of each one in the final sample. 

\begin{compactitem}
    \item \emph{Bias\_diff = 80}: This serves as the reference value for comparing \emph{Bias\_diff\_off} and \emph{Bias\_diff\_on}, essentially representing the initial bias level.
    \item \emph{Bias\_diff\_on = 130}: This sets the ON contrast threshold, defining the increase in brightness required for a pixel to trigger an ON event.
    \item \emph{Bias\_diff\_off = 35}: This sets the OFF contrast threshold, defining the decrease in brightness required for a pixel to trigger an OFF event.
    \item \emph{Bias\_fo = 74}: This adjusts the filtering of rapidly fluctuating light, determining the maximum rate of change in illumination that can be detected.
    \item \emph{Bias\_hpf = 0}: This adjusts the filtering of slow illumination changes, determining the minimum rate of change required for a pixel to output an event.
    \item \emph{Bias\_refr = 68}: This sets the duration for which a pixel remains unresponsive following each event.
\end{compactitem}
As previously noted, it is important to consider that these biases were selected for our specific environment and should be adapted to test the attack on other scenarios. 
Even though they do not influence the attack pipeline in any way, careful adaptation is required to obtain a well-performing clean base model.


\end{document}